\documentclass[letterpaper,twocolumn,aps,prl,floatfix,superscriptaddress,noeprint]{revtex4-2}

\usepackage{float}
\usepackage[caption=false]{subfig}
\usepackage[section]{placeins}
\usepackage{graphicx}	
\usepackage{url}
\usepackage{amsmath}
\usepackage{mathtools}
\usepackage{enumitem}
\usepackage{amssymb}
\usepackage{dsfont}
\usepackage{breqn}
\usepackage{physics}
\usepackage{hyperref}
\hypersetup{colorlinks=true,linkcolor=blue,citecolor=green,filecolor=magenta,urlcolor=blue}
\usepackage[capitalise]{cleveref}

\makeatletter
\let\cat@comma@active\@empty
\makeatother

\graphicspath{ {./figs/}}

\newcommand{\Hhat}{\hat{\mathcal{H}}}

\newcommand{\sigmax}{\hat{\sigma}_x}
\newcommand{\sigmay}{\hat{\sigma}_y}
\newcommand{\sigmaz}{\hat{\sigma}_z}

\begin{document}	
\title{
	Frustration-induced anomalous transport and strong photon decay in waveguide QED
}
\date{\today}
\author{Ron Belyansky}
\email{rbelyans@umd.edu}
\author{Seth Whitsitt}
\author{Rex Lundgren}
\affiliation{Joint Center for Quantum Information and Computer Science, NIST/University of Maryland, College Park, Maryland 20742 USA}
\affiliation{Joint Quantum Institute, NIST/University of Maryland, College Park, Maryland 20742 USA}
\author{Yidan Wang}
\affiliation{Joint Quantum Institute, NIST/University of Maryland, College Park, Maryland 20742 USA}
\author{Andrei Vrajitoarea}
\author{Andrew A. Houck}
\affiliation{Department of Electrical Engineering, Princeton University, Princeton, NJ, USA
}
\author{Alexey V. Gorshkov}
\affiliation{Joint Center for Quantum Information and Computer Science, NIST/University of Maryland, College Park, Maryland 20742 USA}
\affiliation{Joint Quantum Institute, NIST/University of Maryland, College Park, Maryland 20742 USA}
\begin{abstract}
	We study the propagation of photons in a one-dimensional environment consisting of two non-interacting species of photons frustratingly coupled to a single spin-1/2.
	The ultrastrong frustrated coupling leads to an extreme mixing of the light and matter degrees of freedom,  resulting in the disintegration of the spin and a breakdown of the ``dressed-spin", or polaron, description.
	Using a combination of numerical and analytical methods, we show that the elastic response becomes increasingly weak at the effective spin frequency,  showing instead an increasingly strong and broadband response at higher energies. 
	We also show that the photons can decay into multiple photons of smaller energies.
	The total probability of these inelastic processes can be as large as the total elastic scattering rate, or half of the total scattering rate, which is as large as it can be.
	The frustrated spin induces strong anisotropic photon-photon interactions that are dominated by inter-species interactions.
	Our results are relevant to state-of-the-art circuit and cavity quantum electrodynamics experiments. 
\end{abstract}
\maketitle
\pagenumbering{arabic}

Photons propagating in one-dimensional environments are a fundamental building block for quantum optics and waveguide quantum electrodynamics (QED).
While interaction among photons is inherently negligible, strong effective interactions can be induced by coupling the light to atoms, or ``impurities".
Such photon-photon interactions are a crucial ingredient in many technologies ranging from quantum communication to quantum computation and metrology \cite{Roy2017,FriskKockum,Forn-Diaz2019,blais2020circuit}.
Even a single two-level-atom (or a spin-1/2) can induce non-trivial behavior, perfectly reflecting photons whose energy matches the two-level gap $\Delta$, while being transparent for other photons \cite{Shen2006,Shen2007a,Chang2007,Astafiev840}.

This picture can be greatly modified when the light-matter coupling is increased to the so-called ultrastrong coupling (USC) regime of waveguide QED \cite{LeHur2012,Peropadre2013,Goldstein2013,Bera2016,Gheeraert2018,Shi2018}.
This regime has been recently of great experimental and theoretical interest \cite{Gu2017b,FriskKockum,Forn-Diaz2019,LeBoite2020,blais2020circuit}, and has been experimentally realized in superconducting quantum circuits \cite{Forn-Diaz2016,Magazzu2018,PuertasMartinez2019,Kuzmin2019,Leger2019}, allowing the exploration of quantum many-body physics with a \emph{single} artificial atom \cite{VojtaImpurity}.
The hallmark feature of USC regime is the breakdown of the rotating-wave approximation and the description of light and matter as separate entities, which must instead be described by hybridized excitations.

Nevertheless, most light-matter systems do admit an intuitive interpretation in terms of quasi-particles whose behavior closely resembles the bare constituents of the system.
For a two-level atom coupled to a 1D continuum, such a hybridized description is given in terms of a ``dressed spin" or a polaron \cite{Emery1971,Silbey1984,Harris1985}.
The strong dressing of the spin by photons leads to a dramatic Lamb shift of the bare spin frequency $\Delta$ to a renormalized value $\Delta_R\ll\Delta$ \cite{Leggett},
the energy of the polaron excitation.
The propagation of photons in the system can be understood in terms of scattering of free photons off the polaron, with the scattering resonance being shifted from $\Delta$ to $\Delta_R$ \cite{Peropadre2013,Shi2018}.
This renormalized frequency emerges as the natural intrinsic energy scale of the system, with all non-trivial physics, such as photon-photon interactions, occurring in the vicinity of $\Delta_R$.
This intuition can be formalized with the well-known variational polaron transformation, which has been widely successful in describing both static and dynamical observables in various spin-boson systems \cite{Nazir2012,Chin2011,Diaz-Camacho2016,He2018,Sanchez-Burillo2019,Shi2018,Paulisch2018,Bera2014,Roman-Roche2020}.

In this Letter, we introduce a regime of light-matter interaction where the dressed-spin quasi-particle description of the combined light-matter system qualitatively breaks down.
This is induced by ultrastrong frustrated interactions between a single two-level atom and two different species of photons in one dimension \cite{CastroNeto2003,Novais2005}. 
We use matrix-product-state (MPS) numerics together with field-theoretical calculations to study the propagation of a single photon in the system.
At weaker couplings, the elastic scattering shows a peaked response at a renormalized value $\Delta_R< \Delta$, consistent with the polaron interpretation. 
However, at larger couplings, this resonance becomes increasingly weak, and instead there is an emergent \emph{increasingly} large and broadband response at large frequencies $\omega > \Delta_R$.
We also find that inelastic processes, where the photon decays into several smaller-energies photons, can be as important or even \emph{dominate} the elastic scattering.
This decay rate does not peak in the vicinity of $\Delta_R$, in contrast to the polaron scenario, but saturates close to its allowed maximum and persists at very high energies, exceeding even the bare gap $\Delta$.
Both the elastic and inelastic results show that the induced photon-photon interactions can be highly anisotropic, being dominated by interactions between photons of different species.

The model we study is closely related to the problem of a spin coupled to two competing Ohmic baths. The ground-state phase diagram and the spin properties in such a system were originally studied in the context of quantum impurities in magnetically ordered backgrounds \cite{Sengupta2000,Zhu2002,Zarand2002,CastroNeto2003,Khveshchenko2004,Novais2005},
where it was observed that the two competing baths result in the preservation of coherence in the spin dynamics, which was named ``quantum frustration of decoherence" \cite{CastroNeto2003,Novais2005}.
Here we are instead interested in the dynamics of the photons.

\begin{figure}[tb]
	\centering
	\includegraphics[width=\linewidth]{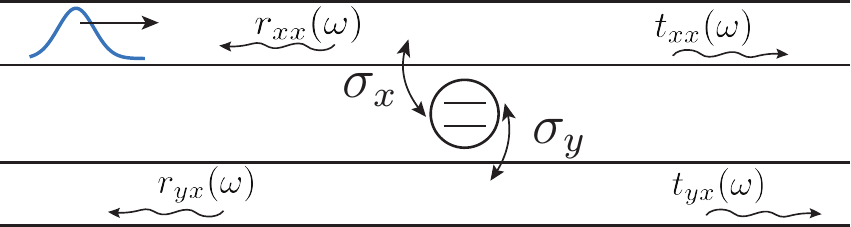}
	\caption{Schematic of the model, where a single spin-1/2 is coupled locally to two independent electromagnetic fields, represented here as two separate waveguides. 
	}
	\label{fig:twobathscat}
\end{figure}
{\it Model.---}We consider a single spin-1/2 that is coupled, via two non-commuting operators, to two species of propagating photons in one-dimension, as shown in \cref{fig:twobathscat}, and as described by the Hamiltonian \cite{CastroNeto2003,Novais2005} ($\hbar=1$)
\begin{dmath}
	\label{eq:real-space-ham}
	\Hhat = \sum_{i=x,y}\int dz \frac{1}{2}\qty(q_i(z)^2+(\partial_z\phi_i(z))^2)-\frac{\Delta}{2}\sigmaz +\pi\sqrt{\alpha_x}q_x(0)\sigmax + \pi\sqrt{\alpha_y}\partial_z\phi_y(0)\sigmay.
\end{dmath}
The two photon species have a linear dispersion $\omega_k=\abs{k}$ and are described by the scalar fields $\phi_i(z)$ satisfying $\comm{q_i(z)}{\phi_j(z')}=-i\delta_{ij}\delta(z-z')$.
Here, $q_i(z)$ and $\phi_i(z)$ could represent the charge and flux degrees of freedom of two superconducting transmission lines \cite{Peropadre2013}, and the spin degree of freedom can be a qubit that is coupled capacitively to one transmission line and inductively to the other \cite{Baksic2014}. 
We note that our results would apply equally well to other geometries, such as a spin coupled to two semi-infinite leads \cite{Goldstein2013}, or a spin 
coupled to two polarizations of a single waveguide as in Ref. \cite{Mahmoodian2019}.
In \cref{eq:real-space-ham}, $\alpha_i$ ($i=x,y$) are the dimensionless coupling constants, which,  for the rest of the Letter, we assume to be equal ($\alpha_x=\alpha_y\equiv \alpha$).

The Hamiltonian in \cref{eq:real-space-ham}
needs to be supplemented with an ultraviolet cutoff $\omega_c$.
The latter can be used to define, via a renormalization group (RG) procedure \cite{Novais2005}, a renormalized spin frequency $\Delta_R$, implicitly given by 
\begin{equation}
	\label{eq:Delta_R}
	\Delta_R = \frac{\Delta}{1+2\alpha\log(\omega_c/\Delta_R)}.
\end{equation}
This quantity, first derived in Refs. \cite{CastroNeto2003,Novais2005}, is close to the bare spin frequency $\Delta$ for small $\alpha\rightarrow 0$, and it decreases as $\alpha$ is increased, approaching $0$ as $\alpha\rightarrow \infty$.
As we show in the next section, for intermediate coupling strengths $\alpha\lesssim0.4$, $\Delta_R$ plays an important role in the photon dynamics, where it can be interpreted as the splitting of the dressed spin, whereas this picture breaks down for larger $\alpha$.

\begin{figure*}[ht!]
	\centering
	\includegraphics[width=\linewidth]{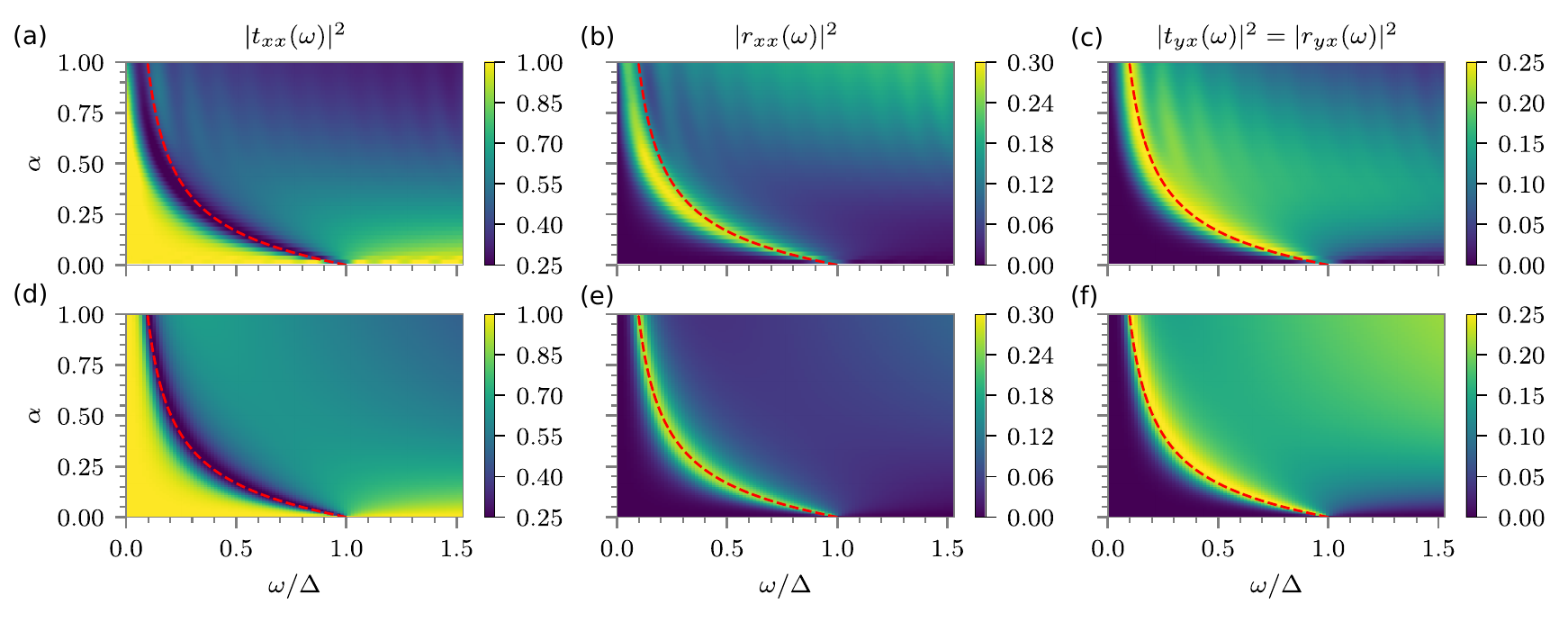}
	\caption{Numerical (a,b,c) and analytical (d,e,f) elastic scattering coefficients corresponding to \cref{fig:twobathscat}, as a function of the incoming frequency $\omega$ and coupling constant $\alpha$. The red dashed line corresponds to $\Delta_R$ from \cref{eq:Delta_R}. The cutoff is $\omega_c=10\Delta$. The oscillating behavior in the numerical plots at large $\alpha$ is a finite-size effect due to the scattering being very broad in space-time.} 
	\label{fig:elas_scat}
\end{figure*}
{\it Anomalous transport.---}We begin by considering the elastic scattering of a single photon. Without loss of generality, we assume an incoming $x$ photon that can scatter elastically in four different ways, as shown in \cref{fig:twobathscat}.
We computed the scattering coefficients both numerically, using an MPS-based approach, and analytically, with diagrammatic perturbation theory.
In order to simulate the system numerically, we use an orthogonal polynomials mapping \cite{Prior,Chin2010} that transforms \cref{eq:real-space-ham} into a one-dimensional tight-binding model with only local interactions (see Supplemental Material \cite{sup}). 
We first use the density matrix renormalization group method to find the ground state of the system and then create a broad-in-frequency single-photon wavepacket on top of it. This state is then evolved for sufficiently long times so that the scattering process has ended. From the resulting state, we extract the elastic probabilities \cite{Sanchez-Burillo2014}, shown in the top row of \cref{fig:elas_scat}, as a function of the incoming frequency $\omega$ and coupling constant $\alpha$.

In order to gain analytical insight into the problem, we use the fact that the elastic S-matrix can be written in terms of spin susceptibilities \cite{Langreth1966,Zarand2004,Fritz2006,Borda2007,Goldstein2013,Bera2016,Magazzu2018}.
For the setup in \cref{fig:twobathscat}, we find \cite{sup}
\begin{equation}\label{eq:refl-trans}
	r_{\alpha\beta}(\omega)  =-i2\pi\alpha\omega\chi_{\alpha\beta}(\omega),\quad
	t_{\alpha\beta}(\omega) = \delta_{\alpha\beta}+r_{\alpha\beta}(\omega),
\end{equation}
where the spin susceptibilities $\chi_{\alpha\beta}(\omega)$ are given by the Fourier-transformed retarded Green's function
\begin{equation}
	\label{eq:spin-susc-def}
	\chi_{\alpha\beta}(\omega)=-\frac{i}{4}\int_{0}^{\infty}dt e^{i\omega t}\expval{\comm{\hat{\sigma}_\alpha(t)}{\hat{\sigma}_\beta(0)}},
\end{equation}
evaluated in the ground state.
\Cref{eq:refl-trans,eq:spin-susc-def} are exact for a single incoming photon, but they can be understood intuitively within linear response formalism.
The scattering of a $\beta$ photon acts as a perturbation $\hat{\sigma}_\beta(0)$ on the spin, and the response $\hat{\sigma}_\alpha(t)$ of the spin describes the emission of an $\alpha$ photon.

The advantage of writing the elastic S-matrix in the form of \cref{eq:refl-trans,eq:spin-susc-def} is that it allows the use of powerful field-theoretical methods.
In particular, we use an Abrikosov psuedo-fermion representation of the spin to perturbatively compute \cref{eq:spin-susc-def} to leading order in $\alpha$, and employ the Dyson equation to sum an infinite subset of diagrams, as in the random-phase-approximation of the Coulomb gas \cite{bruus2004many}. 
We then use the Callan-Symanzik equation together with the RG flow equations from Refs.\ \cite{CastroNeto2003,Novais2005} to improve upon the perturbative results, taking into account the non-perturbative Lamb shift in \cref{eq:Delta_R}.
The end result is (see SM for derivation \cite{sup})
\begin{align}
	\label{eq:xx-susc}
	\chi_{xx}(\omega)=& \frac{(-\Delta+i\pi\alpha\omega)/2}{\Delta^2-\omega^2\qty[\pi^2\alpha^2+\qty(1+2\alpha\log(\frac{\omega_c}{\omega}))^2]-i2\pi\alpha\Delta\omega},\\
	\label{eq:xy-susc}
	\chi_{xy}(\omega)=&\frac{-i\omega(1+2\alpha\log(\omega_c/\omega))/2}{\Delta^2-\omega^2\qty[\pi^2\alpha^2+\qty(1+2\alpha\log(\frac{\omega_c}{\omega}))^2]-i2\pi\alpha\Delta\omega}.
\end{align}
These forms for the susceptibility have a peak near $\Delta_R$ with a width of order $\tau^{-1} \sim \alpha \Delta_R$, where $\tau$ is the lifetime of a spin excitation. 
At small $\alpha$, both expressions reduce to narrows peaks at $\Delta$, since $\Delta_R\rightarrow\Delta$ and $\tau^{-1}\rightarrow 0$ for $\alpha\rightarrow0$.
The resulting transmission and reflection probabilities are shown in the bottom row of \cref{fig:elas_scat}.

We find excellent qualitative agreement between the numerical and analytical results, particularly for $\alpha \lesssim 0.5$.
At very small $\alpha$, we have the standard situation in waveguide QED \cite{Shen2006,Shen2007a,Chang2007,Astafiev840}, where only photons at $\omega \approx \Delta$ are coupled to the spin and experience scattering, being equally split among the four channels in \cref{fig:twobathscat}, and the rest are simply transmitted. 
As $\alpha$ is increased, the location of the resonance drastically decreases, in excellent agreement with the RG predicted $\Delta_R$ in \cref{eq:Delta_R} (red dashed lines in \cref{fig:elas_scat}). 

For $\omega \ll\Delta_R$, \cref{fig:elas_scat} shows perfect transmission for all $\alpha$, indicating that modes with frequencies smaller than $\Delta_R$ are effectively uncoupled from the impurity. This regime is qualitatively similar to that of the usual unfrustrated spin-boson model \cite{Leggett} and the Kondo problem \cite{Emery1974}. 
In the latter, for energies smaller than the Kondo temperature (the equivalent of $\Delta_R$), the impurity is screened and essentially disappears from the problem \cite{Borda2007,Nozieres1974,Emery1974}.

The $\omega >\Delta_R$ regime, on the other hand, is drastically different than in these paradigmatic models and the standard ultrastrong waveguide QED systems  (see SM for a
more detailed comparison to the case when the coupling
operator to both waveguides is the same \cite{sup}). 
Surprisingly, we find that, at large $\alpha$, there is very little transmission, even for $\omega \gg \Delta_R$.
For $\alpha \lesssim 0.4$, the system still admits the effective polaron description, since the strongest elastic response for all scattering channels in \cref{fig:elas_scat} is still concentrated near $\Delta_R$.
This picture changes dramatically for $\alpha \gtrsim 0.4$, where the reflection $\abs{r_{xx}(\omega)}^2$, for example, instead of monotonically decreasing away from the resonance at $\Delta_R$, first decreases but then starts increasing for $\omega > \Delta_R$.
This behavior is more easily seen in the numerical plots but is nonetheless qualitatively consistent with the analytical solution.
In particular, from \cref{eq:xx-susc} we see that, at large $\alpha$ and $\omega\gg\Delta_R$, $\chi_{xx}(\omega)$ decays \emph{sublinearly} $\sim\omega^{-1}\log^{-2}(\omega_c/\omega)$, as was also pointed out in Refs. \cite{CastroNeto2003,Novais2005}. Hence, the reflection coefficient [$\sim \omega\chi_{xx}(\omega)$ from \cref{eq:refl-trans}] \emph{increases}, while the transmission \emph{decreases}, in that regime.
At even higher couplings $\alpha \gtrsim 0.5$, the numerical results show that the $\Delta_R$ resonance in $\abs{r_{xx}(\omega)}^2$ becomes increasingly \emph{weaker},
becoming less intense than the extremely \emph{broadband} response at higher frequencies. 
All this implies that the spectral weight of the spin [$\sim\Im(\chi_{xx}(\omega))$] becomes increasingly spread out over larger energies instead of having a sharp peak at $\Delta_R$.
This anomalous behavior of the elastic reflection and transmission at large $\alpha$, bearing no resemblance to a two-level system, constitutes the first of the two main results of this work.

Another interesting aspect in \cref{fig:elas_scat} is the behavior of the inter-species scattering, $\abs{t_{yx}(\omega)}^2$, where the $\Delta_R$ resonance becomes extremely broad on the $\omega > \Delta_R$ side (note that $\chi_{xy}(\omega)$ [\cref{eq:xy-susc}] approaches a constant for large $\alpha$ and $\omega\gg\Delta_R$). This implies that the incoming $x$ photon can be efficiently converted into a $y$ photon in a wide range of energies.
The inter-species scattering at large $\alpha$ shows significant disagreement between the numerical and analytical results, with the analytics suggesting that $\abs{t_{yx}(\omega)}^2$ increases as $\omega$ is increased away from the $\Delta_R$ resonance. 
The numerics do not show such an increase, but rather show that $\abs{t_{yx}(\omega)}^2$ approaches zero for very large $\omega$ and $\alpha$. 
As we discuss in the next section, the discrepancy in $\abs{t_{yx}(\omega)}^2$ (as well as in $\abs{r_{xx}(\omega)}^2$ and $\abs{t_{xx}(\omega)}^2$) at large $\alpha$ is due to the lack of certain kind of $\order{\alpha^2}$ diagrams in the susceptibility calculation and is related to the presence of substantial inelastic scattering.

{\it Photon decay.---}%
As is well-known, ultrastrong coupling can give rise to number-non-conserving inelastic processes.
The probability of such processes is, however, typically much weaker than the elastic rate and is usually peaked at the vicinity of the polaron energy $\Delta_R$ \cite{Goldstein2013,Zarand2004,Borda2007}. As we now show, these two expectations are strongly invalidated due to the strong frustration in our model,
which constitutes the second main unexpected result of this work.

Conservation of probability implies that  $\abs{t_{xx}(\omega)}^2+\abs{r_{xx}(\omega)}^2 + 2\abs{t_{yx}(\omega)}^2=1-\gamma(\omega)$, where nonzero $\gamma(\omega)$ signifies that the initial $x$ photon of energy $\omega$ can decay into multiple photons of smaller energies.
Direct computation using \cref{eq:xx-susc,eq:xy-susc,eq:refl-trans} yields $\gamma(\omega)=0$, which is certainly incorrect.
In fact, the numerical plots in \cref{fig:elas_scat} show that the total inelastic scattering rate approaches $\approx 0.5$
(for $\omega \gtrsim 0.5\Delta$ and $\alpha \gtrsim 0.6$). 
In those regimes, a photon is therefore as likely to \emph{decay} as to be scattered elastically.
Note that the continuity equation in \cref{eq:refl-trans} constrains that $\max [\gamma]=0.5$, implying that the scattering is nearly \emph{maximally inelastic} in that regime.

\begin{figure}[t]
	\centering
	\includegraphics[width=\linewidth]{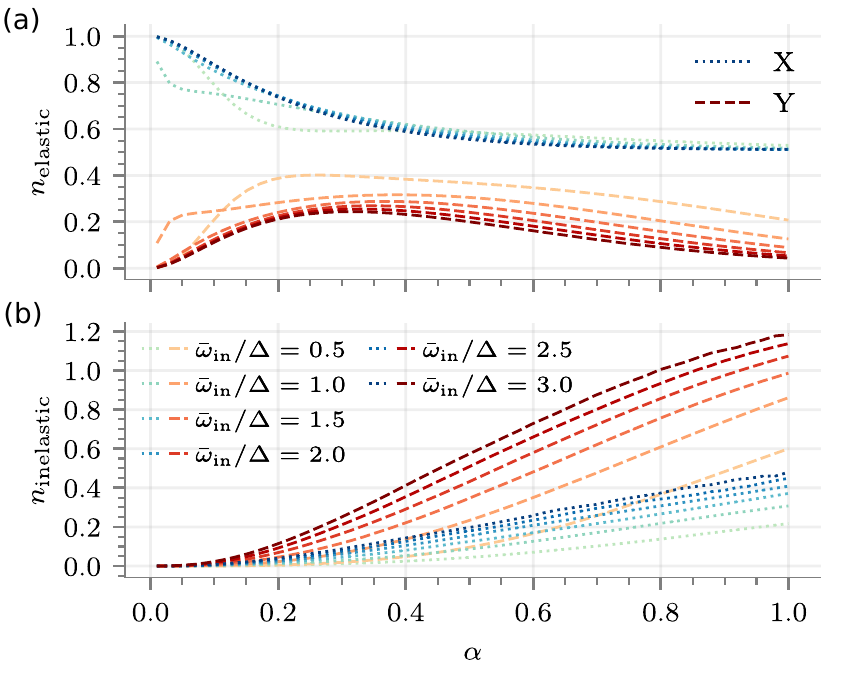}
	\caption{Numerically computed total number of elastic (a) and inelastic (b) particles produced in each waveguide as a function of $\alpha$ for six different incoming wavepackets.  
		The wavepackets are single-particle Gaussians centered at $\bar{\omega}_{in}$ with a standard deviation of $0.2\Delta$. $\omega_c=10\Delta$.}
	\label{fig:inelas_scat}
\end{figure}

To get a deeper understanding of the inelastic scattering, we perform additional numerical simulations and analytical computations.
Numerically, we use narrower wavepackets in order to probe the dependence of the outgoing particles on the energy of the incoming photon.
After the scattering event, we record the total number of elastically and inelastically scattered photons in each waveguide \cite{sup}, shown in \cref{fig:inelas_scat} for six wavepackets with different mean energy.
The first observation from \cref{fig:inelas_scat} 
is that the inelastic emission is highly anisotropic, giving rise to significantly more $y$ photons than $x$ photons, for an initial $x$ wavepacket. Since the scattering process cannot change the state of the spin, the leading-order inelastic process involves four (one incoming and three outgoing) photons and is therefore of order $\alpha^2$  \cite{Goldstein2013}. It is precisely those diagrams which are missing in the susceptibilities in \cref{eq:xx-susc,eq:xy-susc}, explaining also why the analytics become inaccurate for $\alpha \gtrsim 0.5$ where nearly half of the scattering is inelastic.
The four leading-order inelastic processes are $x\rightarrow \{xxx,yyy,xxy,xyy\}$. We denote the probability of these processes by $\gamma_{\mu_1\mu_2\mu_3}(\omega_1,\omega_2,\omega_3;\omega)$ where $\mu_i$ specifies the flavor of the outgoing photon ($x$ or $y$) and $\omega_i$ its frequency. Energy conservation constrains $\omega_1+\omega_2+\omega_3=\omega$. We have computed the leading order diagrams contributing to these processes, and the expressions are provided in the SM \cite{sup}.

We find that the leading-order expression for $\gamma_{xxx}$ exactly matches that \cite{Goldstein2013} of the standard unfrustrated spin-boson model. 
Moreover, three of the four processes are elegantly related to each other to leading order, as follows:
\begin{equation}
	\gamma_{yyy}= \gamma_{xxx}\frac{\omega^2}{\Delta_R^2}\qcomma \gamma_{xxy}=\gamma_{xxx}\frac{\omega_3^2}{\Delta_R^2}.
\end{equation}
The first of these demonstrates that an incoming $x$ photon with energy $\omega>\Delta_R$ is more likely to decay into three $y$ photons, as opposed to three $x$ photons. 
The second relation shows that the $\gamma_{xxy}$ process is more likely to occur than $\gamma_{xxx}$ provided that the energy of the $y$ photon satisfies $\omega_3>\Delta_R$. However, it is far less likely compared to $\gamma_{yyy}$ because if $\omega_3\approx \omega$, energy conservation would require $\omega_1\approx\omega_2\approx0$ and this would highly suppress its probability.
The remaining process, $\gamma_{xyy}$, does not have a simple relation to the other three, but we have verified by direct numerical integration that its total cross-section is of the same order as the one for $\gamma_{xxy}$, and both of these are significantly less important than $\gamma_{yyy}$.
In short, all this demonstrates that, in the regime $\omega\gg\Delta_R$, photons of one flavor decay dominantly to the other.
This agrees qualitatively with \cref{fig:inelas_scat}, since even the smallest energy wavepacket ($\bar{\omega}_{in}/\Delta =0.5$) is in the regime of $\omega>\Delta_R$ for $\alpha\gtrsim 0.2$ (see \cref{fig:elas_scat}).
In fact, for almost all the wavepackets and the range of $\alpha$ in \cref{fig:inelas_scat}, we have $\omega\gg\Delta_R$.

\Cref{fig:inelas_scat} 
also shows that the number of elastic $y$ photons goes to zero at large $\alpha$ and $\omega$, consistent with $t_{yx}(\omega)\rightarrow0$ in that regime, as we discussed in the previous section.
Interestingly, this says that the inter-species scattering can be \emph{completely inelastic}, while also dominating over the intra-species scattering, as we have just shown.
Remarkably, we also see that the number of inelastically produced photons continues to rise as a function of $\bar{\omega}_{in}$, suggesting that $\gamma(\omega)$ remains close to $0.5$ even for $\omega >\Delta\gg \Delta_R$.
This behavior is consistent with the anomalous elastic scattering we identified in the previous section, but we nonetheless expect that $\gamma(\omega)$, as well as the nontrivial elastic scattering [$r_{xx}(\omega),t_{yx}(\omega)$], would eventually decay to zero as $\omega\rightarrow\infty$. While all the presented results are qualitatively independent of the high-energy cutoff, we conjecture that the exact location of this decay may be nonuniversal and may depend on the precise cutoff function for a given physical system.
In the SM \cite{sup}, we compare these results to the situation where the two waveguides in \cref{fig:twobathscat} couple to the spin via the same operator, $\hat{\sigma}_x$, showing that, without the frustrated coupling, the inelastic processes are comparatively negligible and the anomalous elastic transport is absent.

{\it Summary and outlook.---}%
In this work we have shown that ultrastrong frustrated coupling between a two-level system and free photons in 1D leads to novel behavior such as anomalous photon transport and maximal photon decay.
This behavior bears no resemblance to scattering off a two-level system and hence indicates the breakdown of the polaron quasi-particle description.
Instead, this is reminiscent of non-Fermi liquid behavior of quantum impurity models in strongly correlated electron systems \cite{Zarand2004,Borda2007}. 

While in this paper we have focused solely on the equal couplings case $\alpha_x=\alpha_y$, we expect our main results, namely the anomalous transport and strong photon decay, to remain qualitatively valid even in the presence of anisotropic couplings, provided they are both large and similar in magnitude. On the other hand, if the couplings are strongly asymmetric, say $\alpha_x\gg \alpha_y$, the behavior of the system would resemble more the unfrustrated model. The weaker coupling in such a case can be thought of as an unwanted source of dissipation acting on the spin, which would necessarily be present experimentally. Such unwanted dissipation can be similarly included in our model by adding a third waveguide with coupling $\alpha_3\ll \alpha_x\approx \alpha_y$. In superconducting circuits, 
such additional dissipation channels can be made negligible compared to the desired couplings \cite{Forn-Diaz2016,Magazzu2018}, and thus should not qualitatively affect our results.

Future theoretical work can investigate what kind of effective spin-spin interactions as well as novel phases of hybrid light-matter systems can be engineered by adding multiple impurities.
The numerical and analytical methods developed in this work can also be immediately applied in a variety of other situations, such as photons with more exotic dispersions. 
It would also be interesting to develop protocols that make use of the unusual properties of the light-matter system in this work for entanglement generation, single-photon switches and routers, and frequency conversion, among other applications.
Finally, our work may also shed light and inspire future studies on the problems of heat and energy transport, relevant for quantum thermodynamics and quantum chemistry, where similar models to the one studied here appear \cite{Yao2015,Duan2020}.

\begin{acknowledgments}
	We acknowledge valuable discussions with P. Bienias, M. Plenio, I. Boettcher.
	We thank P. Bienias for comments on the manuscript.
	R.B., R.L., Y.W., and A.V.G.\ acknowledge funding by ARO MURI, U.S.~Department of Energy Award No.~DE-SC0019449, NSF PFC at JQI,  DoE ASCR Quantum Testbed Pathfinder program (award No.\ DE-SC0019040), DoE ASCR Accelerated Research in Quantum Computing program (award No.\  DE-SC0020312), NSF PFCQC program, AFOSR, ARL CDQI, and AFOSR MURI. A.V.\ and A.A.H.\ acknowledge funding by NSF (PHY-1607160).
	R.B. acknowledges support of NSERC and FRQNT of Canada. The authors acknowledge the University of Maryland supercomputing resources (\url{http://hpcc.umd.edu}) made available for conducting the research reported in this Letter.
	Numerical simulations were performed using the
	ITensor Library \cite{ITensor}.
\end{acknowledgments}


\begin{thebibliography}{60}%
	\makeatletter
	\providecommand \@ifxundefined [1]{%
		\@ifx{#1\undefined}
	}%
	\providecommand \@ifnum [1]{%
		\ifnum #1\expandafter \@firstoftwo
		\else \expandafter \@secondoftwo
		\fi
	}%
	\providecommand \@ifx [1]{%
		\ifx #1\expandafter \@firstoftwo
		\else \expandafter \@secondoftwo
		\fi
	}%
	\providecommand \natexlab [1]{#1}%
	\providecommand \enquote  [1]{``#1''}%
	\providecommand \bibnamefont  [1]{#1}%
	\providecommand \bibfnamefont [1]{#1}%
	\providecommand \citenamefont [1]{#1}%
	\providecommand \href@noop [0]{\@secondoftwo}%
	\providecommand \href [0]{\begingroup \@sanitize@url \@href}%
	\providecommand \@href[1]{\@@startlink{#1}\@@href}%
	\providecommand \@@href[1]{\endgroup#1\@@endlink}%
	\providecommand \@sanitize@url [0]{\catcode `\\12\catcode `\$12\catcode
		`\&12\catcode `\#12\catcode `\^12\catcode `\_12\catcode `\%12\relax}%
	\providecommand \@@startlink[1]{}%
	\providecommand \@@endlink[0]{}%
	\providecommand \url  [0]{\begingroup\@sanitize@url \@url }%
	\providecommand \@url [1]{\endgroup\@href {#1}{\urlprefix }}%
	\providecommand \urlprefix  [0]{URL }%
	\providecommand \Eprint [0]{\href }%
	\providecommand \doibase [0]{https://doi.org/}%
	\providecommand \selectlanguage [0]{\@gobble}%
	\providecommand \bibinfo  [0]{\@secondoftwo}%
	\providecommand \bibfield  [0]{\@secondoftwo}%
	\providecommand \translation [1]{[#1]}%
	\providecommand \BibitemOpen [0]{}%
	\providecommand \bibitemStop [0]{}%
	\providecommand \bibitemNoStop [0]{.\EOS\space}%
	\providecommand \EOS [0]{\spacefactor3000\relax}%
	\providecommand \BibitemShut  [1]{\csname bibitem#1\endcsname}%
	\let\auto@bib@innerbib\@empty
	\bibitem [{\citenamefont {Roy}\ \emph {et~al.}(2017)\citenamefont {Roy},
		\citenamefont {Wilson},\ and\ \citenamefont {Firstenberg}}]{Roy2017}%
	\BibitemOpen
	\bibfield  {author} {\bibinfo {author} {\bibfnamefont {D.}~\bibnamefont
			{Roy}}, \bibinfo {author} {\bibfnamefont {C.~M.}\ \bibnamefont {Wilson}},\
		and\ \bibinfo {author} {\bibfnamefont {O.}~\bibnamefont {Firstenberg}},\
	}\bibfield  {title} {\bibinfo {title} {{Colloquium : Strongly interacting
				photons in one-dimensional continuum}},\ }\href
	{https://doi.org/10.1103/RevModPhys.89.021001} {\bibfield  {journal}
		{\bibinfo  {journal} {Rev. Mod. Phys.}\ }\textbf {\bibinfo {volume} {89}},\
		\bibinfo {pages} {021001} (\bibinfo {year} {2017})}\BibitemShut {NoStop}%
	\bibitem [{\citenamefont {{Frisk Kockum}}\ \emph {et~al.}(2019)\citenamefont
		{{Frisk Kockum}}, \citenamefont {Miranowicz}, \citenamefont {{De Liberato}},
		\citenamefont {Savasta},\ and\ \citenamefont {Nori}}]{FriskKockum}%
	\BibitemOpen
	\bibfield  {author} {\bibinfo {author} {\bibfnamefont {A.}~\bibnamefont
			{{Frisk Kockum}}}, \bibinfo {author} {\bibfnamefont {A.}~\bibnamefont
			{Miranowicz}}, \bibinfo {author} {\bibfnamefont {S.}~\bibnamefont {{De
					Liberato}}}, \bibinfo {author} {\bibfnamefont {S.}~\bibnamefont {Savasta}},\
		and\ \bibinfo {author} {\bibfnamefont {F.}~\bibnamefont {Nori}},\ }\bibfield
	{title} {\bibinfo {title} {{Ultrastrong coupling between light and matter}},\
	}\href {https://doi.org/10.1038/s42254-018-0006-2} {\bibfield  {journal}
		{\bibinfo  {journal} {Nat. Rev. Phys.}\ }\textbf {\bibinfo {volume} {1}},\
		\bibinfo {pages} {19} (\bibinfo {year} {2019})}\BibitemShut {NoStop}%
	\bibitem [{\citenamefont {Forn-D{\'{i}}az}\ \emph {et~al.}(2019)\citenamefont
		{Forn-D{\'{i}}az}, \citenamefont {Lamata}, \citenamefont {Rico},
		\citenamefont {Kono},\ and\ \citenamefont {Solano}}]{Forn-Diaz2019}%
	\BibitemOpen
	\bibfield  {author} {\bibinfo {author} {\bibfnamefont {P.}~\bibnamefont
			{Forn-D{\'{i}}az}}, \bibinfo {author} {\bibfnamefont {L.}~\bibnamefont
			{Lamata}}, \bibinfo {author} {\bibfnamefont {E.}~\bibnamefont {Rico}},
		\bibinfo {author} {\bibfnamefont {J.}~\bibnamefont {Kono}},\ and\ \bibinfo
		{author} {\bibfnamefont {E.}~\bibnamefont {Solano}},\ }\bibfield  {title}
	{\bibinfo {title} {{Ultrastrong coupling regimes of light-matter
				interaction}},\ }\href {https://doi.org/10.1103/RevModPhys.91.025005}
	{\bibfield  {journal} {\bibinfo  {journal} {Rev. Mod. Phys}\ }\textbf
		{\bibinfo {volume} {91}},\ \bibinfo {pages} {25005} (\bibinfo {year}
		{2019})}\BibitemShut {NoStop}%
	\bibitem [{\citenamefont {Blais}\ \emph {et~al.}(2020)\citenamefont {Blais},
		\citenamefont {Grimsmo}, \citenamefont {Girvin},\ and\ \citenamefont
		{Wallraff}}]{blais2020circuit}%
	\BibitemOpen
	\bibfield  {author} {\bibinfo {author} {\bibfnamefont {A.}~\bibnamefont
			{Blais}}, \bibinfo {author} {\bibfnamefont {A.~L.}\ \bibnamefont {Grimsmo}},
		\bibinfo {author} {\bibfnamefont {S.~M.}\ \bibnamefont {Girvin}},\ and\
		\bibinfo {author} {\bibfnamefont {A.}~\bibnamefont {Wallraff}},\ }\href@noop
	{} {\bibinfo {title} {Circuit quantum electrodynamics}} (\bibinfo {year}
	{2020}),\ \Eprint {https://arxiv.org/abs/2005.12667} {arXiv:2005.12667
		[quant-ph]} \BibitemShut {NoStop}%
	\bibitem [{\citenamefont {Shen}\ and\ \citenamefont {Fan}(2005)}]{Shen2006}%
	\BibitemOpen
	\bibfield  {author} {\bibinfo {author} {\bibfnamefont {J.-T.}\ \bibnamefont
			{Shen}}\ and\ \bibinfo {author} {\bibfnamefont {S.}~\bibnamefont {Fan}},\
	}\bibfield  {title} {\bibinfo {title} {{Coherent Single Photon Transport in a
				One-Dimensional Waveguide Coupled with Superconducting Quantum Bits}},\
	}\href {https://doi.org/10.1103/PhysRevLett.95.213001} {\bibfield  {journal}
		{\bibinfo  {journal} {Phys. Rev. Lett.}\ }\textbf {\bibinfo {volume} {95}},\
		\bibinfo {pages} {213001} (\bibinfo {year} {2005})}\BibitemShut {NoStop}%
	\bibitem [{\citenamefont {Shen}\ and\ \citenamefont {Fan}(2007)}]{Shen2007a}%
	\BibitemOpen
	\bibfield  {author} {\bibinfo {author} {\bibfnamefont {J.-T.}\ \bibnamefont
			{Shen}}\ and\ \bibinfo {author} {\bibfnamefont {S.}~\bibnamefont {Fan}},\
	}\bibfield  {title} {\bibinfo {title} {{Strongly correlated multiparticle
				transport in one dimension through a quantum impurity}},\ }\href
	{https://doi.org/10.1103/PhysRevA.76.062709} {\bibfield  {journal} {\bibinfo
			{journal} {Phys. Rev. A}\ }\textbf {\bibinfo {volume} {76}},\ \bibinfo
		{pages} {062709} (\bibinfo {year} {2007})}\BibitemShut {NoStop}%
	\bibitem [{\citenamefont {Chang}\ \emph {et~al.}(2007)\citenamefont {Chang},
		\citenamefont {S{\o}rensen}, \citenamefont {Demler},\ and\ \citenamefont
		{Lukin}}]{Chang2007}%
	\BibitemOpen
	\bibfield  {author} {\bibinfo {author} {\bibfnamefont {D.~E.}\ \bibnamefont
			{Chang}}, \bibinfo {author} {\bibfnamefont {A.~S.}\ \bibnamefont
			{S{\o}rensen}}, \bibinfo {author} {\bibfnamefont {E.~A.}\ \bibnamefont
			{Demler}},\ and\ \bibinfo {author} {\bibfnamefont {M.~D.}\ \bibnamefont
			{Lukin}},\ }\bibfield  {title} {\bibinfo {title} {{A single-photon transistor
				using nanoscale surface plasmons}},\ }\href
	{https://doi.org/10.1038/nphys708} {\bibfield  {journal} {\bibinfo  {journal}
			{Nat. Phys.}\ }\textbf {\bibinfo {volume} {3}},\ \bibinfo {pages} {807}
		(\bibinfo {year} {2007})}\BibitemShut {NoStop}%
	\bibitem [{\citenamefont {Astafiev}\ \emph {et~al.}(2010)\citenamefont
		{Astafiev}, \citenamefont {Zagoskin}, \citenamefont {Abdumalikov},
		\citenamefont {Pashkin}, \citenamefont {Yamamoto}, \citenamefont {Inomata},
		\citenamefont {Nakamura},\ and\ \citenamefont {Tsai}}]{Astafiev840}%
	\BibitemOpen
	\bibfield  {author} {\bibinfo {author} {\bibfnamefont {O.}~\bibnamefont
			{Astafiev}}, \bibinfo {author} {\bibfnamefont {A.~M.}\ \bibnamefont
			{Zagoskin}}, \bibinfo {author} {\bibfnamefont {A.~A.}\ \bibnamefont
			{Abdumalikov}}, \bibinfo {author} {\bibfnamefont {Y.~A.}\ \bibnamefont
			{Pashkin}}, \bibinfo {author} {\bibfnamefont {T.}~\bibnamefont {Yamamoto}},
		\bibinfo {author} {\bibfnamefont {K.}~\bibnamefont {Inomata}}, \bibinfo
		{author} {\bibfnamefont {Y.}~\bibnamefont {Nakamura}},\ and\ \bibinfo
		{author} {\bibfnamefont {J.~S.}\ \bibnamefont {Tsai}},\ }\bibfield  {title}
	{\bibinfo {title} {Resonance fluorescence of a single artificial atom},\
	}\href {https://doi.org/10.1126/science.1181918} {\bibfield  {journal}
		{\bibinfo  {journal} {Science}\ }\textbf {\bibinfo {volume} {327}},\ \bibinfo
		{pages} {840} (\bibinfo {year} {2010})}\BibitemShut {NoStop}%
	\bibitem [{\citenamefont {{Le Hur}}(2012)}]{LeHur2012}%
	\BibitemOpen
	\bibfield  {author} {\bibinfo {author} {\bibfnamefont {K.}~\bibnamefont {{Le
					Hur}}},\ }\bibfield  {title} {\bibinfo {title} {{Kondo resonance of a
				microwave photon}},\ }\href {https://doi.org/10.1103/PhysRevB.85.140506}
	{\bibfield  {journal} {\bibinfo  {journal} {Phys. Rev. B}\ }\textbf {\bibinfo
			{volume} {85}},\ \bibinfo {pages} {140506} (\bibinfo {year}
		{2012})}\BibitemShut {NoStop}%
	\bibitem [{\citenamefont {Peropadre}\ \emph {et~al.}(2013)\citenamefont
		{Peropadre}, \citenamefont {Zueco}, \citenamefont {Porras},\ and\
		\citenamefont {Garc{\'{i}}a-Ripoll}}]{Peropadre2013}%
	\BibitemOpen
	\bibfield  {author} {\bibinfo {author} {\bibfnamefont {B.}~\bibnamefont
			{Peropadre}}, \bibinfo {author} {\bibfnamefont {D.}~\bibnamefont {Zueco}},
		\bibinfo {author} {\bibfnamefont {D.}~\bibnamefont {Porras}},\ and\ \bibinfo
		{author} {\bibfnamefont {J.~J.}\ \bibnamefont {Garc{\'{i}}a-Ripoll}},\
	}\bibfield  {title} {\bibinfo {title} {{Nonequilibrium and Nonperturbative
				Dynamics of Ultrastrong Coupling in Open Lines}},\ }\href
	{https://doi.org/10.1103/PhysRevLett.111.243602} {\bibfield  {journal}
		{\bibinfo  {journal} {Phys. Rev. Lett.}\ }\textbf {\bibinfo {volume} {111}},\
		\bibinfo {pages} {243602} (\bibinfo {year} {2013})}\BibitemShut {NoStop}%
	\bibitem [{\citenamefont {Goldstein}\ \emph {et~al.}(2013)\citenamefont
		{Goldstein}, \citenamefont {Devoret}, \citenamefont {Houzet},\ and\
		\citenamefont {Glazman}}]{Goldstein2013}%
	\BibitemOpen
	\bibfield  {author} {\bibinfo {author} {\bibfnamefont {M.}~\bibnamefont
			{Goldstein}}, \bibinfo {author} {\bibfnamefont {M.~H.}\ \bibnamefont
			{Devoret}}, \bibinfo {author} {\bibfnamefont {M.}~\bibnamefont {Houzet}},\
		and\ \bibinfo {author} {\bibfnamefont {L.~I.}\ \bibnamefont {Glazman}},\
	}\bibfield  {title} {\bibinfo {title} {{Inelastic Microwave Photon Scattering
				off a Quantum Impurity in a Josephson-Junction Array}},\ }\href
	{https://doi.org/10.1103/PhysRevLett.110.017002} {\bibfield  {journal}
		{\bibinfo  {journal} {Phys. Rev. Lett.}\ }\textbf {\bibinfo {volume} {110}},\
		\bibinfo {pages} {017002} (\bibinfo {year} {2013})}\BibitemShut {NoStop}%
	\bibitem [{\citenamefont {Bera}\ \emph {et~al.}(2016)\citenamefont {Bera},
		\citenamefont {Baranger},\ and\ \citenamefont {Florens}}]{Bera2016}%
	\BibitemOpen
	\bibfield  {author} {\bibinfo {author} {\bibfnamefont {S.}~\bibnamefont
			{Bera}}, \bibinfo {author} {\bibfnamefont {H.~U.}\ \bibnamefont {Baranger}},\
		and\ \bibinfo {author} {\bibfnamefont {S.}~\bibnamefont {Florens}},\
	}\bibfield  {title} {\bibinfo {title} {{Dynamics of a qubit in a
				high-impedance transmission line from a bath perspective}},\ }\href
	{https://doi.org/10.1103/PhysRevA.93.033847} {\bibfield  {journal} {\bibinfo
			{journal} {Phys. Rev. A}\ }\textbf {\bibinfo {volume} {93}},\ \bibinfo
		{pages} {33847} (\bibinfo {year} {2016})}\BibitemShut {NoStop}%
	\bibitem [{\citenamefont {Gheeraert}\ \emph {et~al.}(2018)\citenamefont
		{Gheeraert}, \citenamefont {Zhang}, \citenamefont {S{\'{e}}pulcre},
		\citenamefont {Bera}, \citenamefont {Roch}, \citenamefont {Baranger},\ and\
		\citenamefont {Florens}}]{Gheeraert2018}%
	\BibitemOpen
	\bibfield  {author} {\bibinfo {author} {\bibfnamefont {N.}~\bibnamefont
			{Gheeraert}}, \bibinfo {author} {\bibfnamefont {X.~H.~H.}\ \bibnamefont
			{Zhang}}, \bibinfo {author} {\bibfnamefont {T.}~\bibnamefont
			{S{\'{e}}pulcre}}, \bibinfo {author} {\bibfnamefont {S.}~\bibnamefont
			{Bera}}, \bibinfo {author} {\bibfnamefont {N.}~\bibnamefont {Roch}}, \bibinfo
		{author} {\bibfnamefont {H.~U.}\ \bibnamefont {Baranger}},\ and\ \bibinfo
		{author} {\bibfnamefont {S.}~\bibnamefont {Florens}},\ }\bibfield  {title}
	{\bibinfo {title} {{Particle production in ultrastrong-coupling waveguide
				QED}},\ }\href {https://doi.org/10.1103/PhysRevA.98.043816} {\bibfield
		{journal} {\bibinfo  {journal} {Phys. Rev. A}\ }\textbf {\bibinfo {volume}
			{98}},\ \bibinfo {pages} {043816} (\bibinfo {year} {2018})}\BibitemShut
	{NoStop}%
	\bibitem [{\citenamefont {Shi}\ \emph {et~al.}(2018)\citenamefont {Shi},
		\citenamefont {Chang},\ and\ \citenamefont {Garc{\'{i}}a-Ripoll}}]{Shi2018}%
	\BibitemOpen
	\bibfield  {author} {\bibinfo {author} {\bibfnamefont {T.}~\bibnamefont
			{Shi}}, \bibinfo {author} {\bibfnamefont {Y.}~\bibnamefont {Chang}},\ and\
		\bibinfo {author} {\bibfnamefont {J.~J.}\ \bibnamefont
			{Garc{\'{i}}a-Ripoll}},\ }\bibfield  {title} {\bibinfo {title} {{Ultrastrong
				Coupling Few-Photon Scattering Theory}},\ }\href
	{https://doi.org/10.1103/PhysRevLett.120.153602} {\bibfield  {journal}
		{\bibinfo  {journal} {Phys. Rev. Lett.}\ }\textbf {\bibinfo {volume} {120}},\
		\bibinfo {pages} {153602} (\bibinfo {year} {2018})}\BibitemShut {NoStop}%
	\bibitem [{\citenamefont {Gu}\ \emph {et~al.}(2017)\citenamefont {Gu},
		\citenamefont {Kockum}, \citenamefont {Miranowicz}, \citenamefont {Liu},\
		and\ \citenamefont {Nori}}]{Gu2017b}%
	\BibitemOpen
	\bibfield  {author} {\bibinfo {author} {\bibfnamefont {X.}~\bibnamefont
			{Gu}}, \bibinfo {author} {\bibfnamefont {A.~F.}\ \bibnamefont {Kockum}},
		\bibinfo {author} {\bibfnamefont {A.}~\bibnamefont {Miranowicz}}, \bibinfo
		{author} {\bibfnamefont {Y.-x.}\ \bibnamefont {Liu}},\ and\ \bibinfo {author}
		{\bibfnamefont {F.}~\bibnamefont {Nori}},\ }\bibfield  {title} {\bibinfo
		{title} {{Microwave photonics with superconducting quantum circuits}},\
	}\href {https://doi.org/10.1016/j.physrep.2017.10.002} {\bibfield  {journal}
		{\bibinfo  {journal} {Phys. Rep.}\ }\textbf {\bibinfo {volume} {718-719}},\
		\bibinfo {pages} {1} (\bibinfo {year} {2017})}\BibitemShut {NoStop}%
	\bibitem [{\citenamefont {{Le Boit{\'{e}}}}(2020)}]{LeBoite2020}%
	\BibitemOpen
	\bibfield  {author} {\bibinfo {author} {\bibfnamefont {A.}~\bibnamefont {{Le
					Boit{\'{e}}}}},\ }\bibfield  {title} {\bibinfo {title} {{Theoretical Methods
				for Ultrastrong Light–Matter Interactions}},\ }\href
	{https://doi.org/10.1002/qute.201900140} {\bibfield  {journal} {\bibinfo
			{journal} {Adv. Quantum Technol.}\ }\textbf {\bibinfo {volume} {3}},\
		\bibinfo {pages} {1900140} (\bibinfo {year} {2020})}\BibitemShut {NoStop}%
	\bibitem [{\citenamefont {Forn-D{\'{i}}az}\ \emph {et~al.}(2017)\citenamefont
		{Forn-D{\'{i}}az}, \citenamefont {Garc{\'{i}}a-Ripoll}, \citenamefont
		{Peropadre}, \citenamefont {Orgiazzi}, \citenamefont {Yurtalan},
		\citenamefont {Belyansky}, \citenamefont {Wilson},\ and\ \citenamefont
		{Lupascu}}]{Forn-Diaz2016}%
	\BibitemOpen
	\bibfield  {author} {\bibinfo {author} {\bibfnamefont {P.}~\bibnamefont
			{Forn-D{\'{i}}az}}, \bibinfo {author} {\bibfnamefont {J.~J.}\ \bibnamefont
			{Garc{\'{i}}a-Ripoll}}, \bibinfo {author} {\bibfnamefont {B.}~\bibnamefont
			{Peropadre}}, \bibinfo {author} {\bibfnamefont {J.~L.}\ \bibnamefont
			{Orgiazzi}}, \bibinfo {author} {\bibfnamefont {M.~A.}\ \bibnamefont
			{Yurtalan}}, \bibinfo {author} {\bibfnamefont {R.}~\bibnamefont {Belyansky}},
		\bibinfo {author} {\bibfnamefont {C.~M.}\ \bibnamefont {Wilson}},\ and\
		\bibinfo {author} {\bibfnamefont {A.}~\bibnamefont {Lupascu}},\ }\bibfield
	{title} {\bibinfo {title} {{Ultrastrong coupling of a single artificial atom
				to an electromagnetic continuum in the nonperturbative regime}},\ }\href
	{https://doi.org/10.1038/nphys3905} {\bibfield  {journal} {\bibinfo
			{journal} {Nat. Phys.}\ }\textbf {\bibinfo {volume} {13}},\ \bibinfo {pages}
		{39} (\bibinfo {year} {2017})}\BibitemShut {NoStop}%
	\bibitem [{\citenamefont {Magazz{\`{u}}}\ \emph {et~al.}(2018)\citenamefont
		{Magazz{\`{u}}}, \citenamefont {Forn-D{\'{i}}az}, \citenamefont {Belyansky},
		\citenamefont {Orgiazzi}, \citenamefont {Yurtalan}, \citenamefont {Otto},
		\citenamefont {Lupascu}, \citenamefont {Wilson},\ and\ \citenamefont
		{Grifoni}}]{Magazzu2018}%
	\BibitemOpen
	\bibfield  {author} {\bibinfo {author} {\bibfnamefont {L.}~\bibnamefont
			{Magazz{\`{u}}}}, \bibinfo {author} {\bibfnamefont {P.}~\bibnamefont
			{Forn-D{\'{i}}az}}, \bibinfo {author} {\bibfnamefont {R.}~\bibnamefont
			{Belyansky}}, \bibinfo {author} {\bibfnamefont {J.~L.}\ \bibnamefont
			{Orgiazzi}}, \bibinfo {author} {\bibfnamefont {M.~A.}\ \bibnamefont
			{Yurtalan}}, \bibinfo {author} {\bibfnamefont {M.~R.}\ \bibnamefont {Otto}},
		\bibinfo {author} {\bibfnamefont {A.}~\bibnamefont {Lupascu}}, \bibinfo
		{author} {\bibfnamefont {C.~M.}\ \bibnamefont {Wilson}},\ and\ \bibinfo
		{author} {\bibfnamefont {M.}~\bibnamefont {Grifoni}},\ }\bibfield  {title}
	{\bibinfo {title} {{Probing the strongly driven spin-boson model in a
				superconducting quantum circuit}},\ }\href
	{https://doi.org/10.1038/s41467-018-03626-w} {\bibfield  {journal} {\bibinfo
			{journal} {Nat. Commun.}\ }\textbf {\bibinfo {volume} {9}},\ \bibinfo {pages}
		{1403} (\bibinfo {year} {2018})}\BibitemShut {NoStop}%
	\bibitem [{\citenamefont {{Puertas Mart{\'{i}}nez}}\ \emph
		{et~al.}(2019)\citenamefont {{Puertas Mart{\'{i}}nez}}, \citenamefont
		{L{\'{e}}ger}, \citenamefont {Gheeraert}, \citenamefont {Dassonneville},
		\citenamefont {Planat}, \citenamefont {Foroughi}, \citenamefont {Krupko},
		\citenamefont {Buisson}, \citenamefont {Naud}, \citenamefont
		{Hasch-Guichard}, \citenamefont {Florens}, \citenamefont {Snyman},\ and\
		\citenamefont {Roch}}]{PuertasMartinez2019}%
	\BibitemOpen
	\bibfield  {author} {\bibinfo {author} {\bibfnamefont {J.}~\bibnamefont
			{{Puertas Mart{\'{i}}nez}}}, \bibinfo {author} {\bibfnamefont
			{S.}~\bibnamefont {L{\'{e}}ger}}, \bibinfo {author} {\bibfnamefont
			{N.}~\bibnamefont {Gheeraert}}, \bibinfo {author} {\bibfnamefont
			{R.}~\bibnamefont {Dassonneville}}, \bibinfo {author} {\bibfnamefont
			{L.}~\bibnamefont {Planat}}, \bibinfo {author} {\bibfnamefont
			{F.}~\bibnamefont {Foroughi}}, \bibinfo {author} {\bibfnamefont
			{Y.}~\bibnamefont {Krupko}}, \bibinfo {author} {\bibfnamefont
			{O.}~\bibnamefont {Buisson}}, \bibinfo {author} {\bibfnamefont
			{C.}~\bibnamefont {Naud}}, \bibinfo {author} {\bibfnamefont {W.}~\bibnamefont
			{Hasch-Guichard}}, \bibinfo {author} {\bibfnamefont {S.}~\bibnamefont
			{Florens}}, \bibinfo {author} {\bibfnamefont {I.}~\bibnamefont {Snyman}},\
		and\ \bibinfo {author} {\bibfnamefont {N.}~\bibnamefont {Roch}},\ }\bibfield
	{title} {\bibinfo {title} {{A tunable Josephson platform to explore many-body
				quantum optics in circuit-QED}},\ }\href
	{https://doi.org/10.1038/s41534-018-0104-0} {\bibfield  {journal} {\bibinfo
			{journal} {npj Quantum Inf.}\ }\textbf {\bibinfo {volume} {5}},\ \bibinfo
		{pages} {19} (\bibinfo {year} {2019})}\BibitemShut {NoStop}%
	\bibitem [{\citenamefont {Kuzmin}\ \emph {et~al.}(2019)\citenamefont {Kuzmin},
		\citenamefont {Mehta}, \citenamefont {Grabon}, \citenamefont {Mencia},\ and\
		\citenamefont {Manucharyan}}]{Kuzmin2019}%
	\BibitemOpen
	\bibfield  {author} {\bibinfo {author} {\bibfnamefont {R.}~\bibnamefont
			{Kuzmin}}, \bibinfo {author} {\bibfnamefont {N.}~\bibnamefont {Mehta}},
		\bibinfo {author} {\bibfnamefont {N.}~\bibnamefont {Grabon}}, \bibinfo
		{author} {\bibfnamefont {R.}~\bibnamefont {Mencia}},\ and\ \bibinfo {author}
		{\bibfnamefont {V.~E.}\ \bibnamefont {Manucharyan}},\ }\bibfield  {title}
	{\bibinfo {title} {{Superstrong coupling in circuit quantum
				electrodynamics}},\ }\href {https://doi.org/10.1038/s41534-019-0134-2}
	{\bibfield  {journal} {\bibinfo  {journal} {npj Quantum Inf.}\ }\textbf
		{\bibinfo {volume} {5}},\ \bibinfo {pages} {20} (\bibinfo {year}
		{2019})}\BibitemShut {NoStop}%
	\bibitem [{\citenamefont {L{\'{e}}ger}\ \emph {et~al.}(2019)\citenamefont
		{L{\'{e}}ger}, \citenamefont {Puertas-Mart{\'{i}}nez}, \citenamefont
		{Bharadwaj}, \citenamefont {Dassonneville}, \citenamefont {Delaforce},
		\citenamefont {Foroughi}, \citenamefont {Milchakov}, \citenamefont {Planat},
		\citenamefont {Buisson}, \citenamefont {Naud}, \citenamefont
		{Hasch-Guichard}, \citenamefont {Florens}, \citenamefont {Snyman},\ and\
		\citenamefont {Roch}}]{Leger2019}%
	\BibitemOpen
	\bibfield  {author} {\bibinfo {author} {\bibfnamefont {S.}~\bibnamefont
			{L{\'{e}}ger}}, \bibinfo {author} {\bibfnamefont {J.}~\bibnamefont
			{Puertas-Mart{\'{i}}nez}}, \bibinfo {author} {\bibfnamefont {K.}~\bibnamefont
			{Bharadwaj}}, \bibinfo {author} {\bibfnamefont {R.}~\bibnamefont
			{Dassonneville}}, \bibinfo {author} {\bibfnamefont {J.}~\bibnamefont
			{Delaforce}}, \bibinfo {author} {\bibfnamefont {F.}~\bibnamefont {Foroughi}},
		\bibinfo {author} {\bibfnamefont {V.}~\bibnamefont {Milchakov}}, \bibinfo
		{author} {\bibfnamefont {L.}~\bibnamefont {Planat}}, \bibinfo {author}
		{\bibfnamefont {O.}~\bibnamefont {Buisson}}, \bibinfo {author} {\bibfnamefont
			{C.}~\bibnamefont {Naud}}, \bibinfo {author} {\bibfnamefont {W.}~\bibnamefont
			{Hasch-Guichard}}, \bibinfo {author} {\bibfnamefont {S.}~\bibnamefont
			{Florens}}, \bibinfo {author} {\bibfnamefont {I.}~\bibnamefont {Snyman}},\
		and\ \bibinfo {author} {\bibfnamefont {N.}~\bibnamefont {Roch}},\ }\bibfield
	{title} {\bibinfo {title} {{Observation of quantum many-body effects due to
				zero point fluctuations in superconducting circuits}},\ }\href
	{https://doi.org/10.1038/s41467-019-13199-x} {\bibfield  {journal} {\bibinfo
			{journal} {Nat. Commun.}\ }\textbf {\bibinfo {volume} {10}},\ \bibinfo
		{pages} {5259} (\bibinfo {year} {2019})}\BibitemShut {NoStop}%
	\bibitem [{\citenamefont {Vojta}(2006)}]{VojtaImpurity}%
	\BibitemOpen
	\bibfield  {author} {\bibinfo {author} {\bibfnamefont {M.}~\bibnamefont
			{Vojta}},\ }\bibfield  {title} {\bibinfo {title} {{Impurity quantum phase
				transitions}},\ }\href {https://doi.org/10.1080/14786430500070396} {\bibfield
		{journal} {\bibinfo  {journal} {Philos. Mag.}\ }\textbf {\bibinfo {volume}
			{86}},\ \bibinfo {pages} {1807} (\bibinfo {year} {2006})}\BibitemShut
	{NoStop}%
	\bibitem [{\citenamefont {Emery}\ and\ \citenamefont
		{Luther}(1971)}]{Emery1971}%
	\BibitemOpen
	\bibfield  {author} {\bibinfo {author} {\bibfnamefont {V.~J.}\ \bibnamefont
			{Emery}}\ and\ \bibinfo {author} {\bibfnamefont {A.}~\bibnamefont {Luther}},\
	}\bibfield  {title} {\bibinfo {title} {{Ground-State Properties in the Kondo
				Problem}},\ }\href {https://doi.org/10.1103/PhysRevLett.26.1547} {\bibfield
		{journal} {\bibinfo  {journal} {Phys. Rev. Lett.}\ }\textbf {\bibinfo
			{volume} {26}},\ \bibinfo {pages} {1547} (\bibinfo {year}
		{1971})}\BibitemShut {NoStop}%
	\bibitem [{\citenamefont {Silbey}\ and\ \citenamefont
		{Harris}(1984)}]{Silbey1984}%
	\BibitemOpen
	\bibfield  {author} {\bibinfo {author} {\bibfnamefont {R.}~\bibnamefont
			{Silbey}}\ and\ \bibinfo {author} {\bibfnamefont {R.~A.}\ \bibnamefont
			{Harris}},\ }\bibfield  {title} {\bibinfo {title} {{Variational calculation
				of the dynamics of a two level system interacting with a bath}},\ }\href
	{https://doi.org/10.1063/1.447055} {\bibfield  {journal} {\bibinfo  {journal}
			{J. Chem. Phys.}\ }\textbf {\bibinfo {volume} {80}},\ \bibinfo {pages} {2615}
		(\bibinfo {year} {1984})}\BibitemShut {NoStop}%
	\bibitem [{\citenamefont {Harris}\ and\ \citenamefont
		{Silbey}(1985)}]{Harris1985}%
	\BibitemOpen
	\bibfield  {author} {\bibinfo {author} {\bibfnamefont {R.~A.}\ \bibnamefont
			{Harris}}\ and\ \bibinfo {author} {\bibfnamefont {R.}~\bibnamefont
			{Silbey}},\ }\bibfield  {title} {\bibinfo {title} {{Variational calculation
				of the tunneling system interacting with a heat bath. II. Dynamics of an
				asymmetric tunneling system}},\ }\href {https://doi.org/10.1063/1.449469}
	{\bibfield  {journal} {\bibinfo  {journal} {J. Chem. Phys.}\ }\textbf
		{\bibinfo {volume} {83}},\ \bibinfo {pages} {1069} (\bibinfo {year}
		{1985})}\BibitemShut {NoStop}%
	\bibitem [{\citenamefont {Leggett}\ \emph {et~al.}(1987)\citenamefont
		{Leggett}, \citenamefont {Chakravarty}, \citenamefont {Dorsey}, \citenamefont
		{Fisher}, \citenamefont {Garg},\ and\ \citenamefont {Zwerger}}]{Leggett}%
	\BibitemOpen
	\bibfield  {author} {\bibinfo {author} {\bibfnamefont {A.~J.}\ \bibnamefont
			{Leggett}}, \bibinfo {author} {\bibfnamefont {S.}~\bibnamefont
			{Chakravarty}}, \bibinfo {author} {\bibfnamefont {A.~T.}\ \bibnamefont
			{Dorsey}}, \bibinfo {author} {\bibfnamefont {M.~P.~A.}\ \bibnamefont
			{Fisher}}, \bibinfo {author} {\bibfnamefont {A.}~\bibnamefont {Garg}},\ and\
		\bibinfo {author} {\bibfnamefont {W.}~\bibnamefont {Zwerger}},\ }\bibfield
	{title} {\bibinfo {title} {{Dynamics of the dissipative two-state system}},\
	}\href {https://doi.org/10.1103/RevModPhys.59.1} {\bibfield  {journal}
		{\bibinfo  {journal} {Rev. Mod. Phys.}\ }\textbf {\bibinfo {volume} {59}},\
		\bibinfo {pages} {1} (\bibinfo {year} {1987})}\BibitemShut {NoStop}%
	\bibitem [{\citenamefont {Nazir}\ \emph {et~al.}(2012)\citenamefont {Nazir},
		\citenamefont {McCutcheon},\ and\ \citenamefont {Chin}}]{Nazir2012}%
	\BibitemOpen
	\bibfield  {author} {\bibinfo {author} {\bibfnamefont {A.}~\bibnamefont
			{Nazir}}, \bibinfo {author} {\bibfnamefont {D.~P.~S.}\ \bibnamefont
			{McCutcheon}},\ and\ \bibinfo {author} {\bibfnamefont {A.~W.}\ \bibnamefont
			{Chin}},\ }\bibfield  {title} {\bibinfo {title} {{Ground state and dynamics
				of the biased dissipative two-state system: Beyond variational polaron
				theory}},\ }\href {https://doi.org/10.1103/PhysRevB.85.224301} {\bibfield
		{journal} {\bibinfo  {journal} {Phys. Rev. B}\ }\textbf {\bibinfo {volume}
			{85}},\ \bibinfo {pages} {224301} (\bibinfo {year} {2012})}\BibitemShut
	{NoStop}%
	\bibitem [{\citenamefont {Chin}\ \emph {et~al.}(2011)\citenamefont {Chin},
		\citenamefont {Prior}, \citenamefont {Huelga},\ and\ \citenamefont
		{Plenio}}]{Chin2011}%
	\BibitemOpen
	\bibfield  {author} {\bibinfo {author} {\bibfnamefont {A.~W.}\ \bibnamefont
			{Chin}}, \bibinfo {author} {\bibfnamefont {J.}~\bibnamefont {Prior}},
		\bibinfo {author} {\bibfnamefont {S.~F.}\ \bibnamefont {Huelga}},\ and\
		\bibinfo {author} {\bibfnamefont {M.~B.}\ \bibnamefont {Plenio}},\ }\bibfield
	{title} {\bibinfo {title} {{Generalized polaron ansatz for the ground state
				of the sub-Ohmic spin-boson model: An analytic theory of the localization
				transition}},\ }\href {https://doi.org/10.1103/PhysRevLett.107.160601}
	{\bibfield  {journal} {\bibinfo  {journal} {Phys. Rev. Lett.}\ }\textbf
		{\bibinfo {volume} {107}},\ \bibinfo {pages} {160601} (\bibinfo {year}
		{2011})}\BibitemShut {NoStop}%
	\bibitem [{\citenamefont {D{\'{i}}az-Camacho}\ \emph
		{et~al.}(2016)\citenamefont {D{\'{i}}az-Camacho}, \citenamefont {Bermudez},\
		and\ \citenamefont {Garc{\'{i}}a-Ripoll}}]{Diaz-Camacho2016}%
	\BibitemOpen
	\bibfield  {author} {\bibinfo {author} {\bibfnamefont {G.}~\bibnamefont
			{D{\'{i}}az-Camacho}}, \bibinfo {author} {\bibfnamefont {A.}~\bibnamefont
			{Bermudez}},\ and\ \bibinfo {author} {\bibfnamefont {J.~J.}\ \bibnamefont
			{Garc{\'{i}}a-Ripoll}},\ }\bibfield  {title} {\bibinfo {title} {{Dynamical
				polaron Ansatz: A theoretical tool for the ultrastrong-coupling regime of
				circuit QED}},\ }\href {https://doi.org/10.1103/PhysRevA.93.043843}
	{\bibfield  {journal} {\bibinfo  {journal} {Phys. Rev. A}\ }\textbf {\bibinfo
			{volume} {93}},\ \bibinfo {pages} {43843} (\bibinfo {year}
		{2016})}\BibitemShut {NoStop}%
	\bibitem [{\citenamefont {He}\ \emph {et~al.}(2018)\citenamefont {He},
		\citenamefont {Duan},\ and\ \citenamefont {Chen}}]{He2018}%
	\BibitemOpen
	\bibfield  {author} {\bibinfo {author} {\bibfnamefont {S.}~\bibnamefont
			{He}}, \bibinfo {author} {\bibfnamefont {L.}~\bibnamefont {Duan}},\ and\
		\bibinfo {author} {\bibfnamefont {Q.~H.}\ \bibnamefont {Chen}},\ }\bibfield
	{title} {\bibinfo {title} {{Improved Silbey-Harris polaron ansatz for the
				spin-boson model}},\ }\href {https://doi.org/10.1103/PhysRevB.97.115157}
	{\bibfield  {journal} {\bibinfo  {journal} {Phys. Rev. B}\ }\textbf {\bibinfo
			{volume} {97}},\ \bibinfo {pages} {115157} (\bibinfo {year}
		{2018})}\BibitemShut {NoStop}%
	\bibitem [{\citenamefont {S\'anchez-Burillo}\ \emph {et~al.}(2019)\citenamefont
		{S\'anchez-Burillo}, \citenamefont {Mart\'{\i}n-Moreno}, \citenamefont
		{Garc\'{\i}a-Ripoll},\ and\ \citenamefont {Zueco}}]{Sanchez-Burillo2019}%
	\BibitemOpen
	\bibfield  {author} {\bibinfo {author} {\bibfnamefont {E.}~\bibnamefont
			{S\'anchez-Burillo}}, \bibinfo {author} {\bibfnamefont {L.}~\bibnamefont
			{Mart\'{\i}n-Moreno}}, \bibinfo {author} {\bibfnamefont {J.~J.}\ \bibnamefont
			{Garc\'{\i}a-Ripoll}},\ and\ \bibinfo {author} {\bibfnamefont
			{D.}~\bibnamefont {Zueco}},\ }\bibfield  {title} {\bibinfo {title} {Single
			photons by quenching the vacuum},\ }\href
	{https://doi.org/10.1103/PhysRevLett.123.013601} {\bibfield  {journal}
		{\bibinfo  {journal} {Phys. Rev. Lett.}\ }\textbf {\bibinfo {volume} {123}},\
		\bibinfo {pages} {013601} (\bibinfo {year} {2019})}\BibitemShut {NoStop}%
	\bibitem [{\citenamefont {Paulisch}\ \emph {et~al.}(2018)\citenamefont
		{Paulisch}, \citenamefont {Shi},\ and\ \citenamefont
		{Garcia-Ripoll}}]{Paulisch2018}%
	\BibitemOpen
	\bibfield  {author} {\bibinfo {author} {\bibfnamefont {V.}~\bibnamefont
			{Paulisch}}, \bibinfo {author} {\bibfnamefont {T.}~\bibnamefont {Shi}},\ and\
		\bibinfo {author} {\bibfnamefont {J.~J.}\ \bibnamefont {Garcia-Ripoll}},\
	}\href@noop {} {\bibinfo {title} {Two-photon scattering in usc regime}}
	(\bibinfo {year} {2018}),\ \Eprint {https://arxiv.org/abs/1810.08439}
	{arXiv:1810.08439 [quant-ph]} \BibitemShut {NoStop}%
	\bibitem [{\citenamefont {Bera}\ \emph {et~al.}(2014)\citenamefont {Bera},
		\citenamefont {Nazir}, \citenamefont {Chin}, \citenamefont {Baranger},\ and\
		\citenamefont {Florens}}]{Bera2014}%
	\BibitemOpen
	\bibfield  {author} {\bibinfo {author} {\bibfnamefont {S.}~\bibnamefont
			{Bera}}, \bibinfo {author} {\bibfnamefont {A.}~\bibnamefont {Nazir}},
		\bibinfo {author} {\bibfnamefont {A.~W.}\ \bibnamefont {Chin}}, \bibinfo
		{author} {\bibfnamefont {H.~U.}\ \bibnamefont {Baranger}},\ and\ \bibinfo
		{author} {\bibfnamefont {S.}~\bibnamefont {Florens}},\ }\bibfield  {title}
	{\bibinfo {title} {{Generalized multipolaron expansion for the spin-boson
				model: Environmental entanglement and the biased two-state system}},\ }\href
	{https://doi.org/10.1103/PhysRevB.90.075110} {\bibfield  {journal} {\bibinfo
			{journal} {Phys. Rev. B}\ }\textbf {\bibinfo {volume} {90}},\ \bibinfo
		{pages} {075110} (\bibinfo {year} {2014})}\BibitemShut {NoStop}%
	\bibitem [{\citenamefont {Rom\'an-Roche}\ \emph {et~al.}(2020)\citenamefont
		{Rom\'an-Roche}, \citenamefont {S\'anchez-Burillo},\ and\ \citenamefont
		{Zueco}}]{Roman-Roche2020}%
	\BibitemOpen
	\bibfield  {author} {\bibinfo {author} {\bibfnamefont {J.}~\bibnamefont
			{Rom\'an-Roche}}, \bibinfo {author} {\bibfnamefont {E.}~\bibnamefont
			{S\'anchez-Burillo}},\ and\ \bibinfo {author} {\bibfnamefont
			{D.}~\bibnamefont {Zueco}},\ }\bibfield  {title} {\bibinfo {title} {Bound
			states in ultrastrong waveguide qed},\ }\href
	{https://doi.org/10.1103/PhysRevA.102.023702} {\bibfield  {journal} {\bibinfo
			{journal} {Phys. Rev. A}\ }\textbf {\bibinfo {volume} {102}},\ \bibinfo
		{pages} {023702} (\bibinfo {year} {2020})}\BibitemShut {NoStop}%
	\bibitem [{\citenamefont {{Castro Neto}}\ \emph {et~al.}(2003)\citenamefont
		{{Castro Neto}}, \citenamefont {Novais}, \citenamefont {Borda}, \citenamefont
		{Zar{\'{a}}nd},\ and\ \citenamefont {Affleck}}]{CastroNeto2003}%
	\BibitemOpen
	\bibfield  {author} {\bibinfo {author} {\bibfnamefont {A.~H.}\ \bibnamefont
			{{Castro Neto}}}, \bibinfo {author} {\bibfnamefont {E.}~\bibnamefont
			{Novais}}, \bibinfo {author} {\bibfnamefont {L.}~\bibnamefont {Borda}},
		\bibinfo {author} {\bibfnamefont {G.}~\bibnamefont {Zar{\'{a}}nd}},\ and\
		\bibinfo {author} {\bibfnamefont {I.}~\bibnamefont {Affleck}},\ }\bibfield
	{title} {\bibinfo {title} {{Quantum Magnetic Impurities in Magnetically
				Ordered Systems}},\ }\href {https://doi.org/10.1103/PhysRevLett.91.096401}
	{\bibfield  {journal} {\bibinfo  {journal} {Phys. Rev. Lett.}\ }\textbf
		{\bibinfo {volume} {91}},\ \bibinfo {pages} {096401} (\bibinfo {year}
		{2003})}\BibitemShut {NoStop}%
	\bibitem [{\citenamefont {Novais}\ \emph {et~al.}(2005)\citenamefont {Novais},
		\citenamefont {{Castro Neto}}, \citenamefont {Borda}, \citenamefont
		{Affleck},\ and\ \citenamefont {Zarand}}]{Novais2005}%
	\BibitemOpen
	\bibfield  {author} {\bibinfo {author} {\bibfnamefont {E.}~\bibnamefont
			{Novais}}, \bibinfo {author} {\bibfnamefont {A.~H.}\ \bibnamefont {{Castro
					Neto}}}, \bibinfo {author} {\bibfnamefont {L.}~\bibnamefont {Borda}},
		\bibinfo {author} {\bibfnamefont {I.}~\bibnamefont {Affleck}},\ and\ \bibinfo
		{author} {\bibfnamefont {G.}~\bibnamefont {Zarand}},\ }\bibfield  {title}
	{\bibinfo {title} {{Frustration of decoherence in open quantum systems}},\
	}\href {https://doi.org/10.1103/PhysRevB.72.014417} {\bibfield  {journal}
		{\bibinfo  {journal} {Phys. Rev. B}\ }\textbf {\bibinfo {volume} {72}},\
		\bibinfo {pages} {014417} (\bibinfo {year} {2005})}\BibitemShut {NoStop}%
	\bibitem [{\citenamefont {Sengupta}(2000)}]{Sengupta2000}%
	\BibitemOpen
	\bibfield  {author} {\bibinfo {author} {\bibfnamefont {A.~M.}\ \bibnamefont
			{Sengupta}},\ }\bibfield  {title} {\bibinfo {title} {{Spin in a fluctuating
				field: The Bose(+Fermi) Kondo models}},\ }\href
	{https://doi.org/10.1103/PhysRevB.61.4041} {\bibfield  {journal} {\bibinfo
			{journal} {Phys. Rev. B}\ }\textbf {\bibinfo {volume} {61}},\ \bibinfo
		{pages} {4041} (\bibinfo {year} {2000})}\BibitemShut {NoStop}%
	\bibitem [{\citenamefont {Zhu}\ and\ \citenamefont {Si}(2002)}]{Zhu2002}%
	\BibitemOpen
	\bibfield  {author} {\bibinfo {author} {\bibfnamefont {L.}~\bibnamefont
			{Zhu}}\ and\ \bibinfo {author} {\bibfnamefont {Q.}~\bibnamefont {Si}},\
	}\bibfield  {title} {\bibinfo {title} {{Critical local-moment fluctuations in
				the Bose-Fermi Kondo model}},\ }\href
	{https://doi.org/10.1103/PhysRevB.66.024426} {\bibfield  {journal} {\bibinfo
			{journal} {Phys. Rev. B}\ }\textbf {\bibinfo {volume} {66}},\ \bibinfo
		{pages} {024426} (\bibinfo {year} {2002})}\BibitemShut {NoStop}%
	\bibitem [{\citenamefont {Zar{\'{a}}nd}\ and\ \citenamefont
		{Demler}(2002)}]{Zarand2002}%
	\BibitemOpen
	\bibfield  {author} {\bibinfo {author} {\bibfnamefont {G.}~\bibnamefont
			{Zar{\'{a}}nd}}\ and\ \bibinfo {author} {\bibfnamefont {E.}~\bibnamefont
			{Demler}},\ }\bibfield  {title} {\bibinfo {title} {{Quantum phase transitions
				in the Bose-Fermi Kondo model}},\ }\href
	{https://doi.org/10.1103/PhysRevB.66.024427} {\bibfield  {journal} {\bibinfo
			{journal} {Phys. Rev. B}\ }\textbf {\bibinfo {volume} {66}},\ \bibinfo
		{pages} {024427} (\bibinfo {year} {2002})}\BibitemShut {NoStop}%
	\bibitem [{\citenamefont {Khveshchenko}(2004)}]{Khveshchenko2004}%
	\BibitemOpen
	\bibfield  {author} {\bibinfo {author} {\bibfnamefont {D.~V.}\ \bibnamefont
			{Khveshchenko}},\ }\bibfield  {title} {\bibinfo {title} {{Quantum impurity
				models of noisy qubits}},\ }\href
	{https://doi.org/10.1103/PhysRevB.69.153311} {\bibfield  {journal} {\bibinfo
			{journal} {Phys. Rev. B}\ }\textbf {\bibinfo {volume} {69}},\ \bibinfo
		{pages} {153311} (\bibinfo {year} {2004})}\BibitemShut {NoStop}%
	\bibitem [{\citenamefont {Baksic}\ and\ \citenamefont
		{Ciuti}(2014)}]{Baksic2014}%
	\BibitemOpen
	\bibfield  {author} {\bibinfo {author} {\bibfnamefont {A.}~\bibnamefont
			{Baksic}}\ and\ \bibinfo {author} {\bibfnamefont {C.}~\bibnamefont {Ciuti}},\
	}\bibfield  {title} {\bibinfo {title} {Controlling discrete and continuous
			symmetries in ``superradiant'' phase transitions with circuit qed systems},\
	}\href {https://doi.org/10.1103/PhysRevLett.112.173601} {\bibfield  {journal}
		{\bibinfo  {journal} {Phys. Rev. Lett.}\ }\textbf {\bibinfo {volume} {112}},\
		\bibinfo {pages} {173601} (\bibinfo {year} {2014})}\BibitemShut {NoStop}%
	\bibitem [{\citenamefont {Mahmoodian}(2019)}]{Mahmoodian2019}%
	\BibitemOpen
	\bibfield  {author} {\bibinfo {author} {\bibfnamefont {S.}~\bibnamefont
			{Mahmoodian}},\ }\bibfield  {title} {\bibinfo {title} {{Chiral Light-Matter
				Interaction beyond the Rotating-Wave Approximation}},\ }\href
	{https://doi.org/10.1103/PhysRevLett.123.133603} {\bibfield  {journal}
		{\bibinfo  {journal} {Phys. Rev. Lett.}\ }\textbf {\bibinfo {volume} {123}},\
		\bibinfo {pages} {133603} (\bibinfo {year} {2019})}\BibitemShut {NoStop}%
	\bibitem [{\citenamefont {Prior}\ \emph {et~al.}(2010)\citenamefont {Prior},
		\citenamefont {Chin}, \citenamefont {Huelga},\ and\ \citenamefont
		{Plenio}}]{Prior}%
	\BibitemOpen
	\bibfield  {author} {\bibinfo {author} {\bibfnamefont {J.}~\bibnamefont
			{Prior}}, \bibinfo {author} {\bibfnamefont {A.~W.}\ \bibnamefont {Chin}},
		\bibinfo {author} {\bibfnamefont {S.~F.}\ \bibnamefont {Huelga}},\ and\
		\bibinfo {author} {\bibfnamefont {M.~B.}\ \bibnamefont {Plenio}},\ }\bibfield
	{title} {\bibinfo {title} {{Efficient simulation of strong
				system-environment interactions}},\ }\href
	{https://doi.org/10.1103/PhysRevLett.105.050404} {\bibfield  {journal}
		{\bibinfo  {journal} {Phys. Rev. Lett.}\ }\textbf {\bibinfo {volume} {105}},\
		\bibinfo {pages} {050404} (\bibinfo {year} {2010})}\BibitemShut {NoStop}%
	\bibitem [{\citenamefont {Chin}\ \emph {et~al.}(2010)\citenamefont {Chin},
		\citenamefont {Rivas}, \citenamefont {Huelga},\ and\ \citenamefont
		{Plenio}}]{Chin2010}%
	\BibitemOpen
	\bibfield  {author} {\bibinfo {author} {\bibfnamefont {A.~W.}\ \bibnamefont
			{Chin}}, \bibinfo {author} {\bibfnamefont {{\'{A}}.}~\bibnamefont {Rivas}},
		\bibinfo {author} {\bibfnamefont {S.~F.}\ \bibnamefont {Huelga}},\ and\
		\bibinfo {author} {\bibfnamefont {M.~B.}\ \bibnamefont {Plenio}},\ }\bibfield
	{title} {\bibinfo {title} {{Exact mapping between system-reservoir quantum
				models and semi-infinite discrete chains using orthogonal polynomials}},\
	}\href {https://doi.org/10.1063/1.3490188} {\bibfield  {journal} {\bibinfo
			{journal} {J. Math. Phys.}\ }\textbf {\bibinfo {volume} {51}},\ \bibinfo
		{pages} {092109} (\bibinfo {year} {2010})}\BibitemShut {NoStop}%
	\bibitem [{sup()}]{sup}%
	\BibitemOpen
	\href@noop {} {}\bibinfo {note} {See Supplemental Material at [URL will be
		inserted by publisher] for details on the numerical simulations, derivation
		of the elastic S-matrix and its relation to the spin susceptibilities,
		perturbative calculation of inelastic scattering, and a comparison with a
		system without frustrated couplings, which includes Refs.
		\cite{Suhl1965,coleman_2015,Srednicki2007,Bruognolo2014}.}\BibitemShut
	{Stop}%
	\bibitem [{\citenamefont {Sanchez-Burillo}\ \emph {et~al.}(2014)\citenamefont
		{Sanchez-Burillo}, \citenamefont {Zueco}, \citenamefont {Garcia-Ripoll},\
		and\ \citenamefont {Martin-Moreno}}]{Sanchez-Burillo2014}%
	\BibitemOpen
	\bibfield  {author} {\bibinfo {author} {\bibfnamefont {E.}~\bibnamefont
			{Sanchez-Burillo}}, \bibinfo {author} {\bibfnamefont {D.}~\bibnamefont
			{Zueco}}, \bibinfo {author} {\bibfnamefont {J.~J.}\ \bibnamefont
			{Garcia-Ripoll}},\ and\ \bibinfo {author} {\bibfnamefont {L.}~\bibnamefont
			{Martin-Moreno}},\ }\bibfield  {title} {\bibinfo {title} {{Scattering in the
				ultrastrong regime: Nonlinear optics with one photon}},\ }\href
	{https://doi.org/10.1103/PhysRevLett.113.263604} {\bibfield  {journal}
		{\bibinfo  {journal} {Phys. Rev. Lett.}\ }\textbf {\bibinfo {volume} {113}},\
		\bibinfo {pages} {263604} (\bibinfo {year} {2014})}\BibitemShut {NoStop}%
	\bibitem [{\citenamefont {Langreth}(1966)}]{Langreth1966}%
	\BibitemOpen
	\bibfield  {author} {\bibinfo {author} {\bibfnamefont {D.~C.}\ \bibnamefont
			{Langreth}},\ }\bibfield  {title} {\bibinfo {title} {{Friedel Sum Rule for
				Anderson's Model of Localized Impurity States}},\ }\href
	{https://doi.org/10.1103/PhysRev.150.516} {\bibfield  {journal} {\bibinfo
			{journal} {Phys. Rev.}\ }\textbf {\bibinfo {volume} {150}},\ \bibinfo {pages}
		{516} (\bibinfo {year} {1966})}\BibitemShut {NoStop}%
	\bibitem [{\citenamefont {Zar{\'{a}}nd}\ \emph {et~al.}(2004)\citenamefont
		{Zar{\'{a}}nd}, \citenamefont {Borda}, \citenamefont {von Delft},\ and\
		\citenamefont {Andrei}}]{Zarand2004}%
	\BibitemOpen
	\bibfield  {author} {\bibinfo {author} {\bibfnamefont {G.}~\bibnamefont
			{Zar{\'{a}}nd}}, \bibinfo {author} {\bibfnamefont {L.}~\bibnamefont {Borda}},
		\bibinfo {author} {\bibfnamefont {J.}~\bibnamefont {von Delft}},\ and\
		\bibinfo {author} {\bibfnamefont {N.}~\bibnamefont {Andrei}},\ }\bibfield
	{title} {\bibinfo {title} {{Theory of Inelastic Scattering from Magnetic
				Impurities}},\ }\href {https://doi.org/10.1103/PhysRevLett.93.107204}
	{\bibfield  {journal} {\bibinfo  {journal} {Phys. Rev. Lett.}\ }\textbf
		{\bibinfo {volume} {93}},\ \bibinfo {pages} {107204} (\bibinfo {year}
		{2004})}\BibitemShut {NoStop}%
	\bibitem [{\citenamefont {Fritz}\ \emph {et~al.}(2006)\citenamefont {Fritz},
		\citenamefont {Florens},\ and\ \citenamefont {Vojta}}]{Fritz2006}%
	\BibitemOpen
	\bibfield  {author} {\bibinfo {author} {\bibfnamefont {L.}~\bibnamefont
			{Fritz}}, \bibinfo {author} {\bibfnamefont {S.}~\bibnamefont {Florens}},\
		and\ \bibinfo {author} {\bibfnamefont {M.}~\bibnamefont {Vojta}},\ }\bibfield
	{title} {\bibinfo {title} {{Universal crossovers and critical dynamics of
				quantum phase transitions: A renormalization group study of the pseudogap
				Kondo problem}},\ }\href {https://doi.org/10.1103/PhysRevB.74.144410}
	{\bibfield  {journal} {\bibinfo  {journal} {Phys. Rev. B}\ }\textbf {\bibinfo
			{volume} {74}},\ \bibinfo {pages} {144410} (\bibinfo {year}
		{2006})}\BibitemShut {NoStop}%
	\bibitem [{\citenamefont {Borda}\ \emph {et~al.}(2007)\citenamefont {Borda},
		\citenamefont {Fritz}, \citenamefont {Andrei},\ and\ \citenamefont
		{Zar{\'{a}}nd}}]{Borda2007}%
	\BibitemOpen
	\bibfield  {author} {\bibinfo {author} {\bibfnamefont {L.}~\bibnamefont
			{Borda}}, \bibinfo {author} {\bibfnamefont {L.}~\bibnamefont {Fritz}},
		\bibinfo {author} {\bibfnamefont {N.}~\bibnamefont {Andrei}},\ and\ \bibinfo
		{author} {\bibfnamefont {G.}~\bibnamefont {Zar{\'{a}}nd}},\ }\bibfield
	{title} {\bibinfo {title} {{Theory of inelastic scattering from quantum
				impurities}},\ }\href {https://doi.org/10.1103/PhysRevB.75.235112} {\bibfield
		{journal} {\bibinfo  {journal} {Phys. Rev. B}\ }\textbf {\bibinfo {volume}
			{75}},\ \bibinfo {pages} {235112} (\bibinfo {year} {2007})}\BibitemShut
	{NoStop}%
	\bibitem [{\citenamefont {Bruus}\ and\ \citenamefont
		{Flensberg}(2004)}]{bruus2004many}%
	\BibitemOpen
	\bibfield  {author} {\bibinfo {author} {\bibfnamefont {H.}~\bibnamefont
			{Bruus}}\ and\ \bibinfo {author} {\bibfnamefont {K.}~\bibnamefont
			{Flensberg}},\ }\href@noop {} {\emph {\bibinfo {title} {Many-body quantum
				theory in condensed matter physics: an introduction}}}\ (\bibinfo
	{publisher} {Oxford university press},\ \bibinfo {year} {2004})\BibitemShut
	{NoStop}%
	\bibitem [{\citenamefont {Emery}\ and\ \citenamefont
		{Luther}(1974)}]{Emery1974}%
	\BibitemOpen
	\bibfield  {author} {\bibinfo {author} {\bibfnamefont {V.~J.}\ \bibnamefont
			{Emery}}\ and\ \bibinfo {author} {\bibfnamefont {A.}~\bibnamefont {Luther}},\
	}\bibfield  {title} {\bibinfo {title} {{Low- temperature properties of the
				Kondo Hamiltonian}},\ }\href {https://doi.org/10.1103/PhysRevB.9.215}
	{\bibfield  {journal} {\bibinfo  {journal} {Phys. Rev. B}\ }\textbf {\bibinfo
			{volume} {9}},\ \bibinfo {pages} {215} (\bibinfo {year} {1974})}\BibitemShut
	{NoStop}%
	\bibitem [{\citenamefont {Nozi{\`{e}}res}(1974)}]{Nozieres1974}%
	\BibitemOpen
	\bibfield  {author} {\bibinfo {author} {\bibfnamefont {P.}~\bibnamefont
			{Nozi{\`{e}}res}},\ }\bibfield  {title} {\bibinfo {title} {{A "fermi-liquid"
				description of the Kondo problem at low temperatures}},\ }\href
	{https://doi.org/10.1007/BF00654541} {\bibfield  {journal} {\bibinfo
			{journal} {J. Low Temp. Phys.}\ }\textbf {\bibinfo {volume} {17}},\ \bibinfo
		{pages} {31} (\bibinfo {year} {1974})}\BibitemShut {NoStop}%
	\bibitem [{\citenamefont {Yao}\ \emph {et~al.}(2015)\citenamefont {Yao},
		\citenamefont {Zhou}, \citenamefont {Prior},\ and\ \citenamefont
		{Zhao}}]{Yao2015}%
	\BibitemOpen
	\bibfield  {author} {\bibinfo {author} {\bibfnamefont {Y.}~\bibnamefont
			{Yao}}, \bibinfo {author} {\bibfnamefont {N.}~\bibnamefont {Zhou}}, \bibinfo
		{author} {\bibfnamefont {J.}~\bibnamefont {Prior}},\ and\ \bibinfo {author}
		{\bibfnamefont {Y.}~\bibnamefont {Zhao}},\ }\bibfield  {title} {\bibinfo
		{title} {{Competition between diagonal and off-diagonal coupling gives rise
				to charge-transfer states in polymeric solar cells}},\ }\href
	{https://doi.org/10.1038/srep14555} {\bibfield  {journal} {\bibinfo
			{journal} {Sci. Rep.}\ }\textbf {\bibinfo {volume} {5}},\ \bibinfo {pages}
		{14555} (\bibinfo {year} {2015})}\BibitemShut {NoStop}%
	\bibitem [{\citenamefont {Duan}\ \emph {et~al.}(2020)\citenamefont {Duan},
		\citenamefont {Hsieh}, \citenamefont {Liu}, \citenamefont {Wu},\ and\
		\citenamefont {Cao}}]{Duan2020}%
	\BibitemOpen
	\bibfield  {author} {\bibinfo {author} {\bibfnamefont {C.}~\bibnamefont
			{Duan}}, \bibinfo {author} {\bibfnamefont {C.-Y.}\ \bibnamefont {Hsieh}},
		\bibinfo {author} {\bibfnamefont {J.}~\bibnamefont {Liu}}, \bibinfo {author}
		{\bibfnamefont {J.}~\bibnamefont {Wu}},\ and\ \bibinfo {author}
		{\bibfnamefont {J.}~\bibnamefont {Cao}},\ }\bibfield  {title} {\bibinfo
		{title} {{Unusual Transport Properties with Noncommutative System–Bath
				Coupling Operators}},\ }\href {https://doi.org/10.1021/acs.jpclett.0c00985}
	{\bibfield  {journal} {\bibinfo  {journal} {J. Phys. Chem. Lett.}\ }\textbf
		{\bibinfo {volume} {11}},\ \bibinfo {pages} {4080} (\bibinfo {year}
		{2020})}\BibitemShut {NoStop}%
	\bibitem [{ITe()}]{ITensor}%
	\BibitemOpen
	\href@noop {} {\bibinfo  {journal} {\mbox{ITensor Library} (version 2.1.1)
			http://itensor.org}\ }\BibitemShut {NoStop}%
	\bibitem [{\citenamefont {Suhl}(1965)}]{Suhl1965}%
	\BibitemOpen
	\bibfield  {journal} {  }\bibfield  {author} {\bibinfo {author} {\bibfnamefont
			{H.}~\bibnamefont {Suhl}},\ }\bibfield  {title} {\bibinfo {title}
		{{Dispersion Theory of the Kondo Effect}},\ }\href
	{https://doi.org/10.1103/PhysRev.138.A515} {\bibfield  {journal} {\bibinfo
			{journal} {Phys. Rev.}\ }\textbf {\bibinfo {volume} {138}},\ \bibinfo {pages}
		{A515} (\bibinfo {year} {1965})}\BibitemShut {NoStop}%
	\bibitem [{\citenamefont {Coleman}(2015)}]{coleman_2015}%
	\BibitemOpen
	\bibfield  {author} {\bibinfo {author} {\bibfnamefont {P.}~\bibnamefont
			{Coleman}},\ }\href {https://doi.org/10.1017/CBO9781139020916} {\emph
		{\bibinfo {title} {Introduction to Many-Body Physics}}}\ (\bibinfo
	{publisher} {Cambridge University Press},\ \bibinfo {year}
	{2015})\BibitemShut {NoStop}%
	\bibitem [{\citenamefont {Srednicki}(2007)}]{Srednicki2007}%
	\BibitemOpen
	\bibfield  {author} {\bibinfo {author} {\bibfnamefont {M.}~\bibnamefont
			{Srednicki}},\ }\href@noop {} {\emph {\bibinfo {title} {{Quantum field
					theory}}}}\ (\bibinfo  {publisher} {Cambridge University Press},\ \bibinfo
	{year} {2007})\BibitemShut {NoStop}%
	\bibitem [{\citenamefont {Bruognolo}\ \emph {et~al.}(2014)\citenamefont
		{Bruognolo}, \citenamefont {Weichselbaum}, \citenamefont {Guo}, \citenamefont
		{von Delft}, \citenamefont {Schneider},\ and\ \citenamefont
		{Vojta}}]{Bruognolo2014}%
	\BibitemOpen
	\bibfield  {author} {\bibinfo {author} {\bibfnamefont {B.}~\bibnamefont
			{Bruognolo}}, \bibinfo {author} {\bibfnamefont {A.}~\bibnamefont
			{Weichselbaum}}, \bibinfo {author} {\bibfnamefont {C.}~\bibnamefont {Guo}},
		\bibinfo {author} {\bibfnamefont {J.}~\bibnamefont {von Delft}}, \bibinfo
		{author} {\bibfnamefont {I.}~\bibnamefont {Schneider}},\ and\ \bibinfo
		{author} {\bibfnamefont {M.}~\bibnamefont {Vojta}},\ }\bibfield  {title}
	{\bibinfo {title} {{Two-bath spin-boson model: Phase diagram and critical
				properties}},\ }\href {https://doi.org/10.1103/PhysRevB.90.245130} {\bibfield
		{journal} {\bibinfo  {journal} {Phys. Rev. B}\ }\textbf {\bibinfo {volume}
			{90}},\ \bibinfo {pages} {245130} (\bibinfo {year} {2014})}\BibitemShut
	{NoStop}%
\end{thebibliography}
\end{document}


\title{Supplemental Materials: Frustration-induced anomalous transport and strong photon decay in waveguide QED}
\author{Ron Belyansky}
\author{Seth Whitsitt}
\author{Rex Lundgren}
\affiliation{Joint Center for Quantum Information and Computer Science, NIST/University of Maryland, College Park, Maryland 20742 USA}
\affiliation{Joint Quantum Institute, NIST/University of Maryland, College Park, Maryland 20742 USA}
\author{Yidan Wang}
\affiliation{Joint Quantum Institute, NIST/University of Maryland, College Park, Maryland 20742 USA}
\author{Andrei Vrajitoarea}
\author{Andrew A. Houck}
\affiliation{Department of Electrical Engineering, Princeton University, Princeton, NJ, USA
}
\author{Alexey V. Gorshkov}
\affiliation{Joint Center for Quantum Information and Computer Science, NIST/University of Maryland, College Park, Maryland 20742 USA}
\affiliation{Joint Quantum Institute, NIST/University of Maryland, College Park, Maryland 20742 USA}
\date{\today}

\maketitle
\onecolumngrid

\renewcommand{\theequation}{S\arabic{equation}}
\renewcommand{\thesubsection}{S\arabic{subsection}}
\renewcommand{\thesubsubsection}{\Alph{subsubsection}}

\renewcommand{\bibnumfmt}[1]{[S#1]}
\renewcommand{\citenumfont}[1]{S#1} 

\pagenumbering{arabic}

\makeatletter
\renewcommand{\thefigure}{S\@arabic\c@figure}
\renewcommand \thetable{S\@arabic\c@table}

\tableofcontents
\section{Renormalized spin frequency}
\label{sec:delta-R}

In this section, we reproduce the derivation \cite{Novais2005} of the crossover energy scale between weak and strong coupling, the equivalent of the Kondo temperature, which we associate with the renormalized spin frequency $\Delta_R$ in Eq.\ (2) in the main text.

We employ the renormalization group (RG) procedure as used in Ref.\ \cite{Novais2005}. Let us denote the cutoff at some energy scale $l$ by $\Lambda(l)$ (such that $\Lambda(0)\equiv \omega_c$ is the original cutoff of the problem). 
The RG procedure consists of integrating out the high-energy modes, and thus redefining the cutoff from $\Lambda$ to $\Lambda -d\Lambda$. This leads to the RG flow equations for the coupling constants $\alpha$ and $h\equiv \Delta/\Lambda$ \cite{Novais2005} (that can also be derived from perturbation theory in $\alpha$)
\begin{align}
	\dv{\alpha}{l} &= -2\alpha^2,\\
	\dv{h}{l} &= (1-2\alpha)h,
\end{align}
where $dl=-d\Lambda/\Lambda$, which implies that $l = \log(\frac{\omega_c}{\Lambda})$.

The equation for $\alpha$ can be readily solved,
\begin{dmath}
	\label{eq:alphal}
	\alpha(l) = \frac{\alpha(0)}{1+2l\alpha(0)}.
\end{dmath}
Plugging \cref{eq:alphal} into the differential equation for $h$ we find
\begin{equation}
	\label{eq:h-l}
	h(l)= \frac{e^l \Delta/\omega_c}{1+2\alpha(0)l}.
\end{equation}

Note that both the cutoff $\Lambda$ and $\Delta$ are decreasing as function of $l$. Equivalently, $h$ increases from its initial value of $h(0)=\Delta/\omega_c$.
Eventually, we have $h(l^*) \approx 1$, which occurs when $\Delta(l^*) \equiv \Delta_R = \Lambda(l^*)$, and hence the RG breaks down. The energy scale corresponding to this is $l^* =\log{\omega_c/\Delta_R}$. Plugging this into \cref{eq:h-l} gives Eq.\ (2) from the main text,
\begin{equation}
	\Delta_R = \frac{\Delta}{1+2\alpha\log(\omega_c/\Delta_R)}.
\end{equation}
This differs from the result presented in Refs.\ \cite{CastroNeto2003,Novais2005} in that we kept the $\Delta_R$ on the right-hand-side (whereas these references approximated it by $\Delta$) as it agrees better with the numerical results.

As we show later in \cref{sec:suscs}, we also reproduce exactly this equation by applying the Callan-Symanzik equation to the bare Green's function of the spin.

\section{Numerical methods \label{sec:numerics}}
In this section, we describe the numerical methods we use to compute the single photon transport properties.

We first write the Hamiltonian (Eq.1 of the main text) in terms of bosonic creation and annihilation operators, as follows,
\begin{dmath}
	\Hhat = -\Delta\frac{\hat{\sigma}_z}{2} +\sum_{i=x,y}\qty[\int_{-k_{max}}^{k_{max}}\omega(k)\ahat_{i,k}^\dagger\ahat_{i,k}^{\dagger} dk +\frac{\hat{\sigma}_i}{2} \int_{-k_{max}}^{k_{max}}g(k)(\ahat_{i,k}^\dagger+\ahat_{i,k}^{\dagger})dk]
\end{dmath}
with $\comm{\ahat_{i,k\vphantom{j,k'}}^{\dagger}}{\ahat_{j,k'}^\dagger}=\delta_{ij}\delta(k-k')$.
We make the transformation 
\begin{equation}
	\label{eq:ABtransformation}
	\begin{split}
		\ahat_{i,k} =& \frac{\Ahat_{i,k}+\Bhat_{i,k}}{\sqrt{2}}, \quad k>0,\\
		\ahat_{i,-k} =& \frac{\Ahat_{i,k}-\Bhat_{i,k}}{\sqrt{2}}, \quad k>0,\\
	\end{split}
\end{equation}
which transforms the Hamiltonian into two commuting parts
\begin{equation}
	\label{eq:ABdecoupledHam-LR}
	\begin{split}
		\Hhat &= \Hhat_{XYSB} + \Hhat_{free},\\
		\Hhat_{XYSB} &= -\Delta\frac{\hat{\sigma}_z}{2}+\sum_{i=x,y}\qty[\int_{0}^{\kmax}\omega(k)\Ahat_{i,k}^\dagger\Ahat_{i,k}^{\dagger}+\frac{\hat{\sigma}_i}{2}\int_0^{\kmax}\tilde{g}_i(k)(\Ahat_{i,k}^{\dagger}+\Ahat_{i,k}^\dagger)],\\
		\Hhat_{free} &= \sum_{i=x,y}\qty[\int_{0}^{\kmax}\omega(k)\Bhat_{i,k}^\dagger\Bhat_{i,k}^{\dagger}],
	\end{split}
\end{equation}
where $\tilde{g}(k) = \sqrt{2}g(k)$.
It is enough therefore to only simulate the dynamics of the $\Hhat_{XYSB}$ Hamiltonian.

Explicitly, to determine the single-particle scattering properties, we perform the following procedure. We create a single-particle Gaussian wavepacket on top of the ground state, with amplitude
$c_k =\mathcal{N}e^{-\frac{(k-k_0)^2}{2\sigma^2}+ikx_0}$, where $\mathcal{N}$ is a normalization so that $\int_{0}^{k_{max}}dk \abs{c_k}^2=1$.
Without loss of generality, we choose this excitation to be of the $x$ type (since for $\alpha_x=\alpha_y$ the Hamiltonian is invariant under $x\leftrightarrow y$ exchange.). We then evolve this state in time, which leads to the scattering of the wavepacket off the spin. At long times after the scattering, we can extract several observables such as the elastic scattering amplitudes and number of elastic and inelastic photons in the final state. 

Denoting the ground state of the full Hamiltonian $\Hhat$ by $\ket{GS}=\ket{GS^{XYSB}}\otimes\ket{0^{free}}$, the initial state is 
\begin{dmath}
	\ket{\psi(0)} = \int_{0}^\kmax dk c_k \ahat_{x,k}^\dagger\ket{GS}=\frac{1}{\sqrt{2}}\int_{0}^\kmax dk c_k \Ahat_{x,k}^\dagger\ket{GS^{XYSB}}\otimes\ket{0^{free}}+\ket{GS^{XYSB}}\otimes\frac{1}{\sqrt{2}}\int_{0}^\kmax dk c_k \Bhat_{x,k}^\dagger\ket{0^{free}}.
\end{dmath}
The time-evolved state is 
\begin{dmath}
	\label{eq:psi_t_LR}
	\ket{\psi(t)}=\frac{1}{\sqrt{2}}\ket{\psi_{XYSB}(t)}\otimes \ket{0^{free}}+e^{-iE_{GS}t}\ket{GS^{XYSB}}\otimes\frac{1}{\sqrt{2}}\int_{0}^\kmax dk c_k e^{-i\omega_kt}\Bhat_{x,k}^\dagger\ket{0^{free}},
\end{dmath}
where $\ket{\psi_{XYSB}(t)}\equiv e^{-i\Hhat_{XYSB}t}\int_{0}^\kmax c_k\Ahat_{x,k}^\dagger\ket{GS^{XYSB}}$ and $E_{GS}$ is the ground state energy.

From this state, we can extract the elastic scattering amplitudes as follows \cite{Sanchez-Burillo2014} (with $t=t_\infty$ sufficiently long so that the scattering event has finished):
\begin{align}
	t_{xx} &= \frac{\bra{GS}\ahat_{x,k}\ket{\psi(t_\infty)}}{c_k} =\frac{1}{2}\bra{GS^{XYSB}}\Ahat_{x,k}\ket{\psi_{XYSB}(t_\infty)}+\frac{1}{2}c_ke^{-i(E_{GS}+\omega_k)t_\infty},\\
	r_{xx} &= \frac{\bra{GS}\ahat_{i,- k}\ket{\psi(t_\infty)}}{c_k} =\frac{1}{2}\bra{GS^{XYSB}}\Ahat_{x,k}\ket{\psi_{XYSB}(t_\infty)}-\frac{1}{2}c_ke^{-i(E_{GS}+\omega_k)t_\infty},\\
	t_{xy} &= \frac{\bra{GS}\ahat_{y,k}\ket{\psi(t_\infty)}}{c_k} =\frac{1}{2}\bra{GS^{XYSB}}\Ahat_{y,k}\ket{\psi_{XYSB}(t_\infty)}.
\end{align}

The number of elastic photons generated by a given wavepacket can be found from the above amplitudes by squaring and integrating over all $k$. 
This gives (for $i=x,y$)
\begin{dmath}
	\label{eq:n_elastic}
	\bar{n}_{i,elastic}=\int_{0}^{\kmax}dk \qty(\abs{\bra{GS}\ahat_{i,k}\ket{\psi(t_\infty)}}^2 + \abs{\bra{GS}\ahat_{i,- k}\ket{\psi(t_\infty)}}^2) =\int_{0}^{\kmax}dk\qty[ \frac{1}{2}\abs{\bra{GS^{XYSB}}\Ahat_{i,k}\ket{\psi_{XYSB}(t_\infty)}}^2+\frac{1}{2}\abs{c_k}^2\delta_{x,i}].
\end{dmath}
The number of inelastic photons is
\begin{dmath}
	\label{eq:n_inelastic}
	\bar{n}_{i,inelastic} = \int_{0}^{k_{max}}dk\frac{1}{2}\bra{\psi_{XYSB}(t_\infty)}\Ahat_{i,k}^\dagger\Ahat_{i,k}\ket{\psi_{XYSB}(t_\infty)}-\int_{0}^{k_{max}}dk\frac{1}{2}\bra{GS^{XYSB}}\Ahat_{i,k}^\dagger\Ahat_{i,k}\ket{GS^{XYSB}}-\int_{0}^{k_{max}}dk\frac{1}{2}\abs{\bra{GS^{XYSB}}\Ahat_{i,k}\ket{\psi_{XYSB}(t_\infty)}}^2.
\end{dmath}
Thus we see that all quantities of interest can be obtained from correlation functions and matrix elements of the states $\ket{GS^{XYSB}}$ and $\ket{\psi_{XYSB}(t_\infty)}$.

\subsection{Orthogonal polynomial mapping}
The Hamiltonian $\Hhat_{XYSB}$ from \cref{eq:ABdecoupledHam-LR} describes a system with very nonlocal interactions.
In order to efficiently simulate it with matrix-product-states, we use the orthogonal polynomial mapping introduced in \cite{Prior,Chin2010}, which maps the Hamiltonian into a tight-binding model with only nearest-neighbor interactions.

Here we summarize the main steps of the mapping. For more details, see Refs.\ \cite{Prior,Chin2010}.
We work with the Hamiltonian from \cref{eq:ABdecoupledHam-LR}, reproduced here 
\begin{dmath}
	\label{eqA:spin-boson}
	\Hhat_{XYSB} = -\Delta\frac{\hat{\sigma}_z}{2}+\sum_{i=x,y}\qty[\int_{0}^{\kmax}\omega(k)\Ahat_{i,k}^\dagger\Ahat_{i,k}+\frac{\hat{\sigma}_i}{2}\int_0^{\kmax}\tilde{g}_i(k)(\Ahat_{i,k}+\Ahat_{i,k}^\dagger)],
\end{dmath}
where $\omega(k)=\omega_c k$, $k_{max}=1$, and $\tilde{g}_i(k) =\sqrt{2\alpha_i\omega_ck}$. The resulting spectral functions are
\begin{equation}
	J_i(\omega) = \pi\sum_k\bar{g}_{i}(k)^2\delta(\omega-\omega(k)) =2\pi\alpha_i\omega \theta(\omega_c-\omega).
\end{equation}
We introduce the unitary transformation
\begin{equation}
	U_{i,n}(k) = \tilde{g}_i(k)p_{i,n}(k)\qc n=0,1,\cdots,
\end{equation}
where $p_{i,n}(k)$ are orthonormal polynomials with respect to the measure $d\mu_i(k)=\tilde{g}_i^2(k) dk$ [i.e.\   $\expval{p_{i,n},p_{j,m}} \equiv \int_{0}^\kmax d\mu_i(k) p_{i,n}(k)p_{j,m}(k)=\delta_{nm}\delta_{ij}$],
and a set of new discrete bosonic modes
\begin{dmath}
	\bhat_{i,n}^\dagger = \int_0^{\kmax}dkU_{i,n}(k)A_{i,k}^\dagger,
\end{dmath}
that satisfy $\comm{\bhat_{i,n}}{\bhat_{j,m}^\dagger}=\delta_{ij}\delta_{n,m}$.

Using the recurrence relations of orthogonal polynomials, one can show that the Hamiltonian in \cref{eqA:spin-boson} can be written as 
\begin{dmath}
	\label{eqA:spin-boson-chain}
	\Hhat_{XYSB} = -\Delta\frac{\hat{\sigma}_z}{2}+\sum_{i=x,y}\frac{\hat{\sigma}_i}{2}\sqrt{\alpha_i\omega_c}(\bhat_{i,0}^\dagger+\bhat_{i,0})+\omega_c\sum_{i=x,y}\sum_{n=0}^\infty\nu_n\bhat_{i,n}^\dagger\bhat_{i,n}+\omega_c\sum_{i=x,y}\sum_{n=0}^\infty\qty[\beta_{n+1}\bhat_{i,n}^\dagger\bhat_{i,n+1}+H.c],
\end{dmath}
which describes two semi-infinite tight-binding bosonic chains that are both coupled to the spin via their first site.
For the Ohmic spectral function, the $p_{i,n}$ polynomials are the Jacobi polynomials, and the on-site energies and hopping coefficients are given by
\begin{equation}
	\begin{split}
		\nu_n &= \frac{1}{2}\qty(1+\frac{1}{(1+2n)(3+2n)}),\\
		\beta_{n+1} &= \frac{1+n}{1+n+2n}\sqrt{\frac{1+n}{2+n}}.
	\end{split}
\end{equation}

Using the inverse transformation 
\begin{equation}
	\Ahat_{i,k}^\dagger = \sum_{n}U_{i,n}(k)\bhat_{i,n}^\dagger,
\end{equation}
we can convert measurements in the $\bhat$ basis to observables in frequency space. 
For example, the frequency-mode occupation is (for $k=k'$)
\begin{dmath}
	\expval{\Ahat_{i,k}^\dagger\Ahat_{i,k'}}=\sum_{n,m=0}^\infty U_{i,n}(k)U_{i,m}(k')\expval{\bhat_{i,n}^\dagger\bhat_{i,m}}.
\end{dmath}
Note that this is an exact mapping, provided the length of the chains is infinite.
In practice, the length of the chains is truncated to a finite value $L$,
and the dimension of each bosonic Hilbert space is truncated to a finite value $d$.
We varied these parameters and found that $L=250$, $d=5$, and bond dimension of $\chi = 30$ are adequate to obtain converging results for the scattering for most values of $\alpha$. 

\section{Elastic S-matrix in terms of spin susceptibilities}
In this section, we derive the relation between the elastic scattering coefficients and the spin susceptibilities, given in Eqs.\ (3,4) in the main text.

Let us write the Hamiltonian as 
\begin{align}
	\Hhat =& \Hhat_0 + \Vhat,\\
	\Hhat_0 =& \sum_k \omega_k\ahat_{x,k}^\dagger\ahat_{x,k}+\sum_k \omega_k\ahat_{y,k}^\dagger\ahat_{y,k},\\
	\Vhat =& -\frac{\Delta}{2}\hat{\sigma}_z+ \frac{\hat{\sigma}_x}{2} \sum_kg_{x,k}(\ahat_{x,k}^\dagger+\ahat_{x,k})+\frac{\hat{\sigma}_y}{2} \sum_kg_{x,k}(\ahat_{y,k}^\dagger+\ahat_{y,k}).
\end{align}

We are interested in the $S$-matrix between a particle with momentum $k$ in bath $i$ and a particle with momentum $k'$ in bath $j$: 
\begin{equation}
	\label{eq:S-matrix-def}
	S_{jk',ik} = \bra*{\psi_{jk'}^-}\ket*{\psi_{ik}^+},
\end{equation}
where $\ket{\psi_{ik}^\pm}$ are the exact incoming and outgoing scattering eigenstates. 
Following Ref.\ \cite{Suhl1965}, we can write these eigenstates as follows:
\begin{equation}
	\ket{\psi_{i k}^\pm} = \ahat_{i,k}^\dagger\ket{\psi_0}+\ket{\chi^\pm_i},
\end{equation}
where $\ket{\psi_0}$ is the ground state of the full $\Hhat$ with energy $E_0$, and $\ket{\chi^{\pm}_i}$ are the states of the scattered particles.
Since $\ket{\psi_{ik}^\pm}$ are eigenstates of $\Hhat$ with energy $E_0+\omega_k$, Schrodinger's equation implies 
\begin{equation}
	\label{eq:schrod-scat-state}
	(\Hhat-\omega_k-E_0)\qty(\ahat_{i,k}^\dagger\ket{\psi_0}+\ket{\chi_i})=0.
\end{equation}
Furthermore,
\begin{equation}
	\label{eq:Hadag-comm}
	\Hhat \ahat_{i,k}^\dagger\ket{\psi_0} = \qty(\ahat_{i,k}^\dagger\Hhat-\comm{\ahat_{i,k}^\dagger}{\Hhat})\ket{\psi_0} = (E_0+\omega_k)\ahat_{i,k}^\dagger\ket{\psi_0} +\frac{\hat{\sigma}_i}{2}g_{i,k}\ket{\psi_0}.
\end{equation}
Substituting this into \cref{eq:schrod-scat-state} gives
\begin{equation}
	(E_0+\omega_k-\Hhat)\ket{\chi_i}=\frac{\hat{\sigma}_i}{2}g_{i,k}\ket{\psi_0} \rightarrow \ket{\chi^\pm_i} = \frac{1}{E_0+\omega_k-\Hhat \pm i\epsilon}\frac{\hat{\sigma}_i}{2}g_{i,k}\ket{\psi_0},
\end{equation}
with $\epsilon >0$ taken to zero at the very end.
From this we find
\begin{equation}
	\ket{\psi_{ik}^{+}}-\ket{\psi_{ik}^-} = -2\pi i \delta(E_0-\omega_k-\Hhat)\frac{\hat{\sigma}_i}{2}g_{i,k}\ket{\psi_0}.
\end{equation}

Plugging this back into the S-matrix \cref{eq:S-matrix-def} gives
\begin{equation}
	S_{jk',ik} = \delta_{kk'}\delta_{ij} - 2\pi i \delta(\omega_{k'}-\omega_k)g_{ik}\bra*{\psi_{jk'}^-}\frac{\hat{\sigma}_i}{2}\ket{\psi_{0}}
	\equiv  \delta_{kk'}\delta_{ij} - 2\pi i \delta(\omega_{k'}-\omega_k)T_{jk',ik},
\end{equation}
where we defined the $T$-matrix
\begin{dmath}
	\label{eq:T-matrix-def}
	T_{jk',ik} = g_{ik}\bra*{\psi_{jk'}^-}\frac{\hat{\sigma}_i}{2}\ket{\psi_{0}} = g_{ik}\bra*{\psi_{0}}\ahat_{jk'}\frac{\hat{\sigma}_i}{2}\ket{\psi_{0}}+g_{ik}g_{jk'}\bra*{\psi_{0}}\frac{\hat{\sigma}_j}{2}\frac{1}{E_0+\omega_k-\Hhat+i\epsilon}\frac{\hat{\sigma}_i}{2}\ket{\psi_{0}}.
\end{dmath}
To evaluate the first term, we perform a similar manipulation as in \cref{eq:Hadag-comm}:
\begin{align}
	\label{eq:Ha-comm}
	\Hhat \ahat_{j,k'}\ket{\psi_0} =& \qty(\ahat_{j,k'}\Hhat-\comm{\ahat_{j,k'}}{\Hhat})\ket{\psi_0} = (E_0-\omega_{k'})\ahat_{j,k'}\ket{\psi_0} -\frac{\hat{\sigma}_j}{2}g_{j,k'}\ket{\psi_0}\\ &\rightarrow
	\ahat_{j,k'}\ket{\psi_0} = \frac{1}{E_0-\omega_{k'}-\Hhat-i\epsilon}\frac{\hat{\sigma}_j}{2}g_{j,k'}\ket{\psi_0}.
\end{align}
Inserting this into \cref{eq:T-matrix-def} yields
\begin{dmath}
	T_{jk',ik} =
	g_{ik}g_{jk'}\qty[\bra*{\psi_{0}}\frac{\hat{\sigma}_j}{2}\frac{1}{E_0+\omega_k-\Hhat+i\epsilon}\frac{\hat{\sigma}_i}{2}\ket{\psi_{0}}
	+
	\bra*{\psi_{0}}\frac{\hat{\sigma}_i}{2}\frac{1}{E_0-\omega_{k'}-\Hhat-i\epsilon}\frac{\hat{\sigma}_j}{2}\ket{\psi_{0}}],
\end{dmath}
which, when $\omega_k=\omega_{k'}$, we recognize as the Fourier transform of the retarded Green's function
\begin{equation}
	T_{jk',ik} = g_{ik}g_{jk'} G_{ji}^R(\omega+i\epsilon)
\end{equation}
with 
\begin{equation}
	\label{eq:spinsusc}
	G_{ji}^R(\omega+i\epsilon) = -\frac{i}{4}\int_0^\infty dt \, e^{i(\omega+i\epsilon)t}\expval{\comm{\hat{\sigma}_j(t)}{\hat{\sigma}_i}},
\end{equation}
which we equivalently refer to as the spin susceptibility in the main text. 

\section{Derivation of the spin susceptibilities}
\label{sec:suscs}

In this section, we explicitly compute the spin susceptibility, Eq.~\eqref{eq:spinsusc}. 
We will do so by first computing the imaginary-time Matsubara Green's function,
\beq
\mathcal{G}_{j i}(i \Omega) = \int_0^{\beta} d \tau \, e^{i \Omega_n \tau} \left\langle \mathcal{T}_{\tau} \hS_j(\tau) \hS_i(0) \right\rangle.
\label{eq:matsG}
\eeq
Here, the imaginary-time dependence of operators is $\hS_j(\tau) = e^{H \tau} \hS_j e^{- H \tau}$, and $\mathcal{T}_{\tau}$ is the imaginary-time ordering operator.
This function may be computed using the usual diagrammatic perturbation theory if we can use Wick's theorem, after which we may obtain the spin susceptibility by analytic continuation \cite{coleman_2015}:
\beq
G_{ji}^R(\omega+i\epsilon) = - \frac{1}{4} \mathcal{G}_{j i}(i \Omega \rightarrow \omega + i \epsilon).
\eeq
However, the Pauli matrices do not satisfy Wick's theorem.
We can get around this by using an Abrikosov pseudo-fermion representation of the spins.
We introduce a two-component set of fermions, $\{ \chi_a , \chi^{\dagger}_b \} = \delta_{ab}$, $a, b = 1,2$, related to the spin operators by
\bea
\hS_x &=& \chi^{\dagger}_1 \chi_2 + \chi^{\dagger}_2 \chi_1, \nn
\hS_y &=& - i \left(\chi^{\dagger}_1 \chi_2 - \chi^{\dagger}_2 \chi_1 \right), \nn
\hS_z &=& \chi^{\dagger}_1 \chi_1 - \chi^{\dagger}_2 \chi_2.
\eea
This is only a faithful representation of the spin operators in the subspace $\chi^{\dagger}_1 \chi_1 + \chi^{\dagger}_2 \chi_2 = 1$.
However, we can project to this subspace using the Popov-Fedotov trick of using an imaginary chemical potential $\mu = -i \pi/2 \beta$, which results in a cancellation between the unphysical subspaces \cite{coleman_2015}.
This method requires us to work at finite temperature during intermediate calculations, but below we will always take the $\beta \rightarrow \infty$ limit as early as possible.

\begin{figure}[htb]
	\centering
	\includegraphics[width=\linewidth]{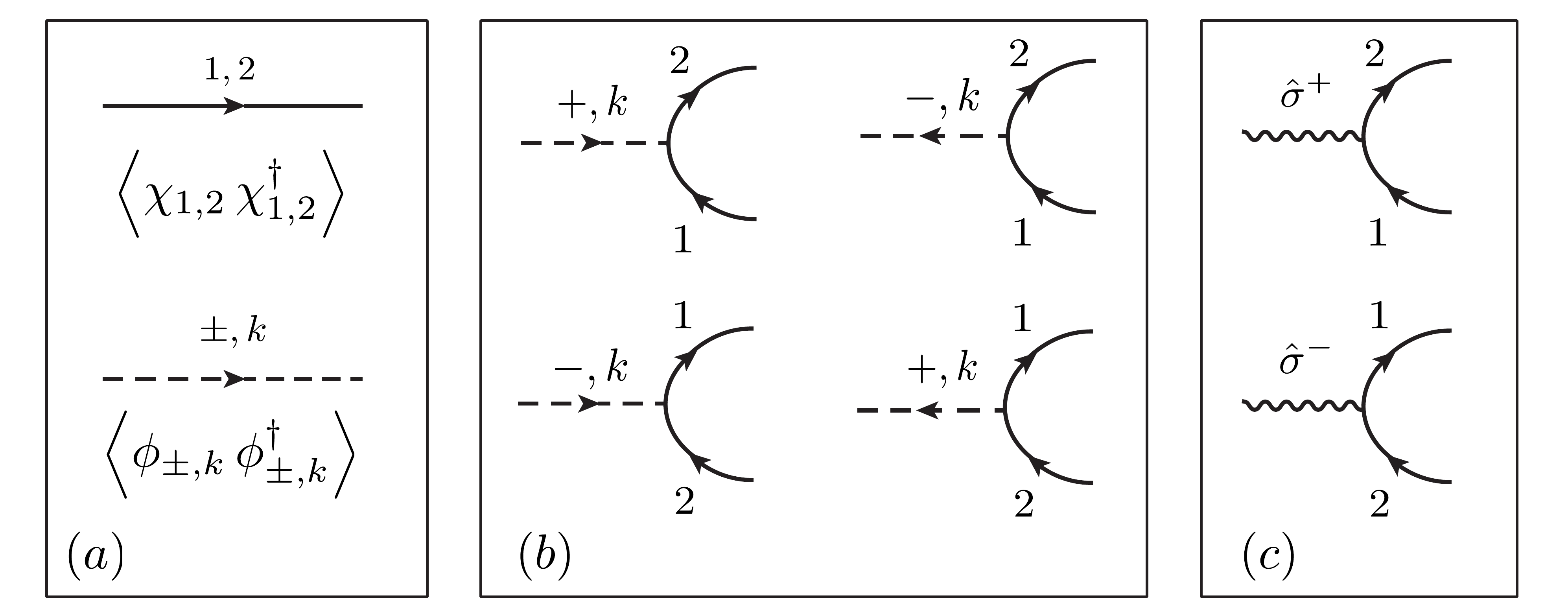}
	\caption{The Feynman rules, which follow from Eq.~\eqref{eq:euclagrangian}. (a) The fermionic (top) and bosonic (bottom) propagators, which are given in Eqs.~\eqref{eq:propi}-\eqref{eq:propf}. (b) The interaction vertices for our theory. Each vertex contributes a factor of $\sum_k g_k/\sqrt{2}$. The dependence of diagrams on the spectral function comes from internal bosonic propagators connecting two such vertices. (c) Diagrammatic representation of the insertion of external spin operators, which are composite when written in terms of the fermionic $\chi_a$ fields.}
	\label{fig:feyn_rules}
\end{figure}
We may now express the system as a coherent-state path integral.
We introduce the bosonic fields $\phi_{\pm,k} = (a_{x,k} \pm i a_{y,k})/\sqrt{2}$, after which our model may be described by the Lagrangian
\bea
\label{eq:euclagrangian}
L &=& \sum_{i = \pm}\sum_k \phi_{i,k}^{\dagger} \left( \partial _{\tau} + \omega_k \right) \phi_{i,k} + \sum_{a,b = 1}^2 \chi^{\dagger}_a \left[ \left( \partial_{\tau} - \mu \right) \delta_{ab} - \frac{\Delta}{2} \sigma^z_{ab} \right] \chi_b \nn
&& + \ \frac{1}{\sqrt{2}} \chi^{\dagger}_1 \chi_2 \sum_k g_k \left( \phi_{-,k} + \phi^{\dagger}_{+,k} \right) + \frac{1}{\sqrt{2}} \chi^{\dagger}_2 \chi_1 \sum_k g_k \left( \phi_{+,k} + \phi^{\dagger}_{-,k} \right).
\eea
In this form, it is straightforward to treat the interactions $g_k$ perturbatively using a Feynman-diagram expansion and the Matsubara formalism. 
We have the bare fermionic Green's functions
\beq
\label{eq:propi}
\delta_{ab} \Pi_a(\tau) = \langle \mathcal{T}_{\tau} \chi_a(\tau) \chi^{\dagger}_b(0) \rangle, \qquad \Pi_a(i \alpha_n) = \int_0^{\beta} d \tau \, e^{i \alpha_n \tau} \Pi_a(\tau),
\eeq
with
\beq
\Pi_1(i \alpha_n) = -\frac{1}{i \alpha_n + \mu + \Delta/2}, \qquad \Pi_2(i \alpha_n) = -\frac{1}{i \alpha_n + \mu - \Delta/2},
\eeq
where $\alpha_n = \pi (2n+1)/\beta$, $n \in \mathbb{Z}$.
Similarly, the bosonic propagators are
\beq
\delta_{ij} \delta_{k k'} D_k(\tau) = \langle \mathcal{T}_{\tau} \phi_{i,k}(\tau) \phi^{\dagger}_{j,k'}(0) \rangle, \qquad D_k(i \Omega_n) = \int_0^{\beta} d \tau \, e^{i \Omega_n \tau} D_k(\tau),
\eeq
with
\beq
\label{eq:propf}
D_k(i \Omega_n) = \frac{1}{- i \Omega_n + \omega_k}
\eeq
and $\Omega_n = 2 \pi n/\beta$, $n \in \mathbb{Z}$.
The interaction terms in Eq.~\eqref{eq:euclagrangian} result in four interaction vertices, which contribute dependence on $g_k$.
Since we are interested in correlation functions of the spin operators, we also introduce a diagrammatic notation representing the composite operators $\hS^+ = \chi^{\dagger}_1 \chi_2$ and $\hS^- = \chi^{\dagger}_2 \chi_1$.
The Feynman rules for this theory are displayed in Fig.~\ref{fig:feyn_rules}.

As a demonstration of this formalism, we obtain the spin susceptibility in the non-interacting ($\alpha=0$) limit by computing the diagrams with a single fermion loop and no bosonic propagators, 
\bea
\label{eq:g0}
\mathcal{G}_{+-}\left( i \Omega_n \right) &=& \int_0^{\beta} d\tau e^{i \Omega_n \tau} \langle \mathcal{T} \chi^{\dagger}_1(\tau) \chi_2(\tau) \chi^{\dagger}_2(0) \chi_1(0) \rangle \nn
&=& -\frac{1}{\beta} \sum_{i \alpha_n} \frac{1}{(i \alpha_n + \mu + \Delta/2)(i \alpha_n + i \Omega_n + \mu - \Delta/2)} \nn
&=& \frac{\tanh\left( \beta \Delta/2 \right)}{\Delta - i \Omega_n}.
\eea
We also have $\mathcal{G}_{-+}(i \Omega_n) = \mathcal{G}_{+-}(-i \Omega_n)$ and $\mathcal{G}_{++} = \mathcal{G}_{--} = 0$.
Going back to the $xy$ basis, taking $\beta = \infty$, and analytically continuing, we obtain the expected form for the spin susceptibilities for $\alpha = 0$:
\bea
\label{eq:freesusc}
G^R_{xx}(\omega + i \epsilon) &=& G^R_{yy}(\omega + i \epsilon) = \frac{\Delta/2}{(\omega + i \epsilon)^2 - \Delta^2}, \nn
G^R_{xy}(\omega + i \epsilon) &=& -G^R_{yx}(\omega + i \epsilon) = \frac{i\omega/2}{(\omega + i \epsilon)^2 - \Delta^2}.
\eea
These expressions could be simply obtained through a direct computation at zero temperature with the spin operators, but the diagrammatic approach is useful for including interactions.

We note that the susceptibilities in Eq.~\eqref{eq:freesusc} have a simple pole located at the bare spin frequency, and such a pole can never be shifted or broadened by computing a finite number of diagrams. Therefore, we will use both the Callan-Symanzik equations and a Dyson equation to sum an infinite number of diagrams, which will result in a change in the analytic structures of the susceptibilities.

As discussed in Sec.~\ref{sec:delta-R}, if we perform an RG transformation on our system, redefining the cutoff from $\omega_c \equiv \Lambda(0)$ to some lower cutoff $\Lambda(l)$, we obtain the flow equations
\begin{align}
	\Lambda\dv{\alpha}{\Lambda} &= 2\alpha^2,\\
	\Lambda\dv{h}{\Lambda} &= -(1-2\alpha)h,
\end{align}
where $h \equiv \Delta/\Lambda$.
In addition to coupling renormalization, it turns out that one also needs to renormalize the spin operators under an RG transformation, and one may show that, perturbatively,
\beq
\sigma_i(l) \approx (1 - \alpha \log \omega_c/\Lambda) \sigma_i(0), \qquad (i = x,y),
\eeq
implying a flow for the operators,
\beq
\Lambda\dv{\sigma_i}{\Lambda} = \alpha \, \sigma_i, \qquad (i = x,y).
\eeq
These flow equations, first obtained in Ref.~\cite{Novais2005} using a Wilsonian momentum-shell RG scheme, may also be obtained by treating Eq.~\eqref{eq:euclagrangian} using the conventional methods of quantum field theory. 
We now use the fact that the susceptibilities should be independent of an RG transformation, $dG^R_{ji}/d\Lambda = 0$.
Taking into account any explicit and implicit dependence on the cutoff, this implies the Callan-Symanzik equation,
\beq
\left[ \Lambda \pdv{\Lambda} - h (1 - 2 \alpha) \pdv{h} + 4 \alpha^2 \pdv{\alpha} + 2 \alpha \right] G^R_{ji} = 0.
\eeq
The general solution to this partial differential equation is
\beq
\label{eq:CSsolution}
G^R_{ji} = \frac{f_{ji}(\bar{h}(\Lambda),\bar{\alpha}(\Lambda))}{\omega + 2 \alpha \omega \log \Lambda/\omega},
\eeq
where the $f_{ji}$ are arbitrary functions of the ``running couplings,'' defined as
\beq
\bar{h}(\Lambda) = \frac{h \Lambda}{\omega + 2 \alpha \omega \log \Lambda/\omega}, \qquad \bar{\alpha}(\Lambda) = \frac{\alpha}{1 + 2 \alpha \log \Lambda/\omega}.
\eeq
Comparing this general solution to the leading-order expressions of Eq.~\eqref{eq:freesusc}, we may read off the $\alpha = 0$ limit of the functions $f_{ji}$, and then use the $\alpha$-dependence implied by the solution of the differential equation to find
\bea
\label{eq:CSimproved}
G^R_{xx} &=& \frac{\Delta/2}{\omega^2 (1 + 2 \alpha \log \omega_c/\omega)^2 - \Delta^2}, \nn
G^R_{xy} &=& \frac{i \omega (1 + 2 \alpha \log \omega_c/\omega)/2}{\omega^2 (1 + 2 \alpha \log \omega_c/\omega)^2 - \Delta^2},
\eea
where we have plugged in $\Lambda(l=0) = \omega_c$ to give expressions in terms of the initial cutoff and the bare quantities, and $\omega$ has a small positive imaginary part.
We see that both expressions no longer diverge at $\omega = \Delta$, but instead they have poles at $\omega = \Delta_R$, where $\Delta_R$ satisfies
\beq
\Delta_R = \frac{\Delta}{1 + 2 \alpha \log(\omega_c/\Delta_R)}.
\eeq
From Eq.~\eqref{eq:CSimproved}, we see that the effect of solving the Callan-Symanzik equation was to sum the ``leading logarithms,'' which are terms of the form $\alpha^n \log^n \omega_c/\omega$ at $n$th order in perturbation theory.

\begin{figure}[htb]
	\centering
	\includegraphics[width=\linewidth]{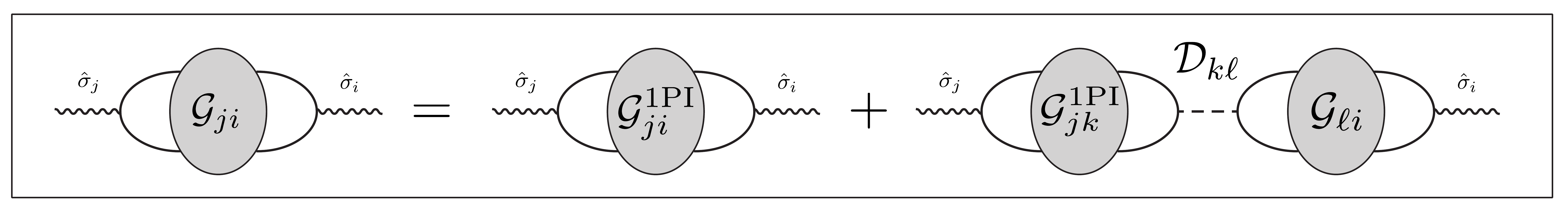}
	\caption{Diagrammatic representation of the Dyson equation, Eq.~\eqref{eq:dysoneq}.}
	\label{fig:dysoneq}
\end{figure}
Although we have succeeded in capturing the renormalization of the spin frequency using the Callan-Symanzik equations, they still predict a sharp behavior at $\omega = \Delta_R$, whereas we expect interactions to broaden the peak near the renormalized spin frequency.
We rectify this by computing an additional infinite set of diagrams using Dyson's equation. 
We first consider the one-particle irreducible Green's functions, $\mathcal{G}^{\mathrm{1PI}}_{ji}$, which are defined to be the complete set of diagrams that cannot be split in two by cutting a single propagator. 
By the structure of the interactions, the only possible propagators which can be cut to disconnect a susceptibility diagram is a bosonic propagator.
As a result, we have the exact relation (in matrix notation)
\beq
\mathcal{G} = \mathcal{G}^{\mathrm{1PI}} + \mathcal{G}^{\mathrm{1PI}} \mathcal{D} \mathcal{G}.
\label{eq:dysoneq}
\eeq
See Fig.~\ref{fig:dysoneq} for a diagrammatic representation of this Dyson equation.
Here, $\mathcal{D}$ is the result from computing the diagrams.
We find the simple structure, $\mathcal{D}^{++} = \mathcal{D}^{--} = 0$ and $\mathcal{D}^{+-} = \mathcal{D}^{-+}$.
An explicit calculation gives
\beq
\mathcal{D}^{+-} = - \pi \alpha |\Omega_n|.
\eeq
Then the full Matsubara Green's function satisfies
\beq
\mathcal{G} = \left[ \mathbb{I} - \mathcal{G}^{\mathrm{1PI}} \mathcal{D} \right]^{-1} \mathcal{G}^{\mathrm{1PI}}.
\eeq

We now approximate the full Green's function by just using the leading-order result, Eq.~\eqref{eq:g0}, for $\mathcal{G}^{\mathrm{1PI}}$.
This corresponds to summing all possible ``bubble diagrams'' which contribute to the susceptibility, which is reminiscent of the RPA approximation in the dense electron gas.
In this approximation, we obtain the susceptibilities as 
\bea
G^R_{xx}(\omega ) &=& \frac{ \left( \Delta  - i \pi \alpha \omega \right)/2}{\left( 1 + \pi^2 \alpha^2 \right)\omega^2 + i 2 \pi \alpha \Delta \omega - \Delta^2}, \nn
G^R_{xy}(\omega) &=& \frac{i \omega/2}{\left( 1 + \pi^2 \alpha^2 \right)\omega^2 + i 2 \pi \alpha \Delta \omega - \Delta^2}.
\label{eq:RPA}
\eea
We see that the inclusion of these diagrams has resulted in a finite imaginary part in the denominator, which removes the pole on the real-$\omega$ axis. 
We may now furthermore use the Callan-Symanzik equation to sum the leading logarithms.
After matching Eq.~\eqref{eq:CSsolution} to Eq.~\eqref{eq:RPA}, we obtain the spin susceptibilities given in Eqs.~(5,6) of the main text.

As noted in the main text, to this order, we have found that the above expressions do not lead to any contribution to inelastic scattering ($\gamma(\omega)$ in the main text). 
We have checked that including all $O(\alpha)$ contributions to $\mathcal{G}^{\mathrm{1PI}}$ still does not lead to inelastic scattering.
We believe that including $O(\alpha^2)$ contributions to $\mathcal{G}^{\mathrm{1PI}}$ will lead to inelastic contributions, which is consistent with \cref{sec:inelastic-scat},  where we show that inelastic contributions to the $S$-matrix appear at $O(\alpha^2)$.

\section{Inelastic scattering}
\label{sec:inelastic-scat}

In this section, we will consider the leading contributions to inelastic scattering in perturbation theory using the diagrammatic approach developed in \cref{sec:suscs}. 
To proceed, we need a relation between time-ordered expectation values and $S$ matrix elements.
Such a relation is called the LSZ reduction formula in relativistic quantum field theory \cite{Srednicki2007}, but we can follow the derivation for our present system and derive a non-relativistic analogue of the reduction formula.
If we consider the scattering of $n$ photons with momenta $k_1$, $k_2$, ..., $k_n$ into a state with $n'$ photons with momenta $k_{1'}$, $k_{2'}$, ..., $k_{n'}$, the $S$ matrix element is given by
\bea
S &=& i^{n + n'} \int dt_{1'} \, e^{i \omega_{k_{1'}} t_{1'}} \left( - i \partial_{t_{1'}} + \omega_{k_{1'}} \right) \cdots \nn
&& \qquad \quad \  \ dt_{1} \, e^{- i \omega_{k_{1}} t_{1}} \left( i \partial_{t_{1}} + \omega_{k_{1}} \right) \cdots \nn
&& \qquad \times  \left\langle \mathcal{T}_t \left\{ \phi_{k_{1'}}(t_{1'})  \cdots  \phi^{\dagger}_{k_{1}}(t_{1})  \cdots \right\} \right\rangle.
\eea
Here, $\omega_k = |k|$ is the energy of the photon. Note that the real time-ordered correlation function appears in this expression, which is related to the Matsubara correlation functions in \cref{sec:suscs} by a Wick rotation. 
This expression greatly simplifies after Fourier transforming to frequency space.
When we evaluate diagrams using Wick's theorem, we will come across the following bosonic contractions from the external legs of Feynman diagrams,
\beq
\left\langle \mathcal{T}_t \left\{ \phi_{k}(t_1) \phi^{\dagger}_k(t) \right\} \right\rangle = -i \int \frac{d \omega}{2 \pi} \frac{e^{- i \omega (t_1 - t)}}{\omega_k - (\omega + i \epsilon)}.
\eeq
These will set the external legs of the Feynman diagrams to their on-shell values, $\omega = \omega_{k}$, and additionally cancel out the contribution of the external propagators.
Then the $S$ matrix elements simply become
\beq
S = \left\langle \mathcal{T}_t \left\{ \phi_{k_{1'}}(\omega_{k_{1'}})  \cdots  \phi^{\dagger}_{k_{1}}(\omega_{k_1})  \cdots \right\} \right\rangle_{\mathrm{amp}.}.
\eeq
That is, we evaluate the diagram in momentum space with on-shell external legs and omit the external propagators (i.e.\ we ``amputate'' the legs). 

\begin{figure}[htb]
	\centering
	\includegraphics[width=\linewidth]{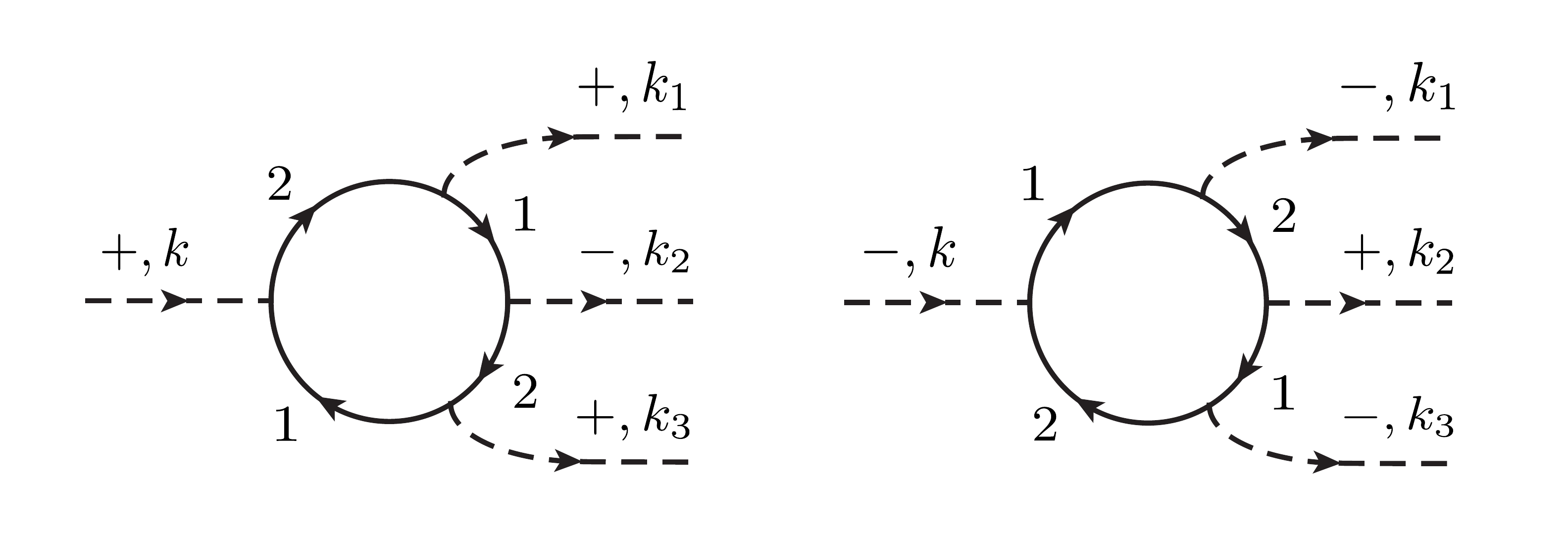}
	\caption{The nonzero diagrams contributing to the scattering of one photon into three photons.}
	\label{fig:inelastdiags}
\end{figure}

The symmetry of the model greatly reduces the number of diagrams we need to consider.
In particular, the Hamiltonian has a U(1) symmetry \cite{Bruognolo2014}, and the conserved charge [which is more easily apparent in terms of the $\pm$ photons $\ahat_{\pm} =\frac{1}{2}(\ahat_x\pm i\ahat_y)$] is
\begin{equation}
	Q = \frac{\hat{\sigma}_z+1}{2} + \sum_{k}(\ahat_{-,k}^\dagger\ahat_{-,k}-\ahat_{+,k}^\dagger\ahat_{+,k}).
\end{equation}
Since the scattering cannot change the spin (as this would require the existence of bound spin-photon eigenstates), it follows that the quantity $\sum_{k}(\ahat_{-,k}^\dagger\ahat_{-,k}-\ahat_{+,k}^\dagger\ahat_{+,k})$ must be conserved in any scattering process.

We now consider the simplest case, where one photon scatters into three photons. The discussion above means that there are only two nonzero diagrams, pictured in Fig.~\ref{fig:inelastdiags}. 
An explicit calculation finds that to lowest order we have
\begin{align}
	\label{eq:ampl1}
	\bra{0} \ahat_{+,k_1}\ahat_{-,k_2}\ahat_{+,k_3}S \ahat_{+,k}^\dagger\ket{0} = \frac{\alpha^2 \omega_c^2}{4}\sqrt{\omega\omega_1\omega_2\omega_3}\frac{2\Delta-\omega_1-\omega_3}{(\omega-\Delta)(\omega_1-\Delta)(\omega_3-\Delta)(\omega_2+\Delta)},\\
	\bra{0} \ahat_{-,k_1}\ahat_{+,k_2}\ahat_{-,k_3}S \ahat_{-,k}^\dagger\ket{0} = \frac{\alpha^2 \omega_c^2}{4}\sqrt{\omega\omega_1\omega_2\omega_3}\frac{2\Delta+\omega_1+\omega_3}{(\omega+\Delta)(\omega_1+\Delta)(\omega_3+\Delta)(\omega_2-\Delta)},
	\label{eq:ampl2}
\end{align}
together with the energy conservation condition $\omega = \omega_1+\omega_2+\omega_3$ (enforced by a delta function, which has been omitted in these expressions). 
Denoting \cref{eq:ampl1} by a function $f(\omega_1,\omega_2,\omega_3;\omega)$, we see that \cref{eq:ampl2} is simply $f(-\omega_1,-\omega_2,-\omega_3;-\omega)$.
Alternatively, each amplitude can be found from the other by substituting $\Delta\rightarrow -\Delta$ (up to an overall minus sign). This is because the two diagrams in \cref{fig:inelastdiags} are related to each other by replacing $+$ photons with $-$ photons and $-$ photons by $+$ photons. From \cref{eq:euclagrangian} we see that this requires flipping the sign of $\Delta$.

Converting \cref{eq:ampl1,eq:ampl2} to the $xy$ basis, we find the four amplitudes
\begin{dmath}
	4\bra{0} \ahat_{x,k_1}\ahat_{x,k_2}\ahat_{x,k_3}S \ahat_{x,k}^\dagger \ket{0}=
	f(\omega_1,\omega_3,\omega_2;\omega)
	+f(\omega_1,\omega_2,\omega_3;
	\omega)
	+f(\omega_2,\omega_1,\omega_3;
	\omega)
	+f(-\omega_2,-\omega_1,-\omega_3;
	-\omega)
	+f(-\omega_1,-\omega_2,-\omega_3;
	-\omega)
	+f(-\omega_1,-\omega_3,-\omega_2;-\omega),
\end{dmath}
\begin{dmath}
	i^{3}4\bra{0} \ahat_{y,k_1}\ahat_{y,k_2}\ahat_{y,k_3}S \ahat_{x,k}^\dagger \ket{0}=
	-f(\omega_1,\omega_3,\omega_2;\omega)
	-f(\omega_1,\omega_2,\omega_3;
	\omega)
	-f(\omega_2,\omega_1,\omega_3;
	\omega)
	+f(-\omega_2,-\omega_1,-\omega_3;
	-\omega)
	+f(-\omega_1,-\omega_2,-\omega_3;
	-\omega)
	+f(-\omega_1,-\omega_3,-\omega_2;-\omega),
\end{dmath}
\begin{dmath}
	i^{2}4\bra{0} \ahat_{x,k_1}\ahat_{y,k_2}\ahat_{y,k_3}S \ahat_{x,k}^\dagger \ket{0}=
	-f(\omega_1,\omega_3,\omega_2;\omega)
	-f(\omega_1,\omega_2,\omega_3;
	\omega)
	+f(\omega_2,\omega_1,\omega_3;
	\omega)
	+f(-\omega_2,-\omega_1,-\omega_3;
	-\omega)
	-f(-\omega_1,-\omega_2,-\omega_3;
	-\omega)
	-f(-\omega_1,-\omega_3,-\omega_2;-\omega),
\end{dmath}
\begin{dmath}
	i4\bra{0} \ahat_{x,k_1}\ahat_{x,k_2}\ahat_{y,k_3}S \ahat_{x,k}^\dagger \ket{0}=
	-f(\omega_1,\omega_3,\omega_2;\omega)
	+f(\omega_1,\omega_2,\omega_3;
	\omega)
	+f(\omega_2,\omega_1,\omega_3;
	\omega)
	-f(-\omega_2,-\omega_1,-\omega_3;
	-\omega)
	-f(-\omega_1,-\omega_2,-\omega_3;
	-\omega)
	+f(-\omega_1,-\omega_3,-\omega_2;-\omega).
\end{dmath}
The resulting probabilities are: for $x \rightarrow xxx$ 
\begin{align}
	\gamma_{xxx}(\omega_1,\omega_2,\omega_3;\omega) =&
	\frac{\alpha^4 \omega_c^4}{16}\omega_1\omega_2\omega_3\omega\abs{ \Delta }^2\times\\ \nonumber
	&\abs{\frac{ 3 \Delta ^4-\Delta ^2 \left(\omega^2-\omega_1\omega_2 -\omega_2\omega_3-\omega_1\omega_2\right)-\omega \omega_1 \omega_2\omega_3}{(\omega-\Delta ) (\Delta +\omega) (\Delta -\omega_1) (\Delta +\omega_1) (\Delta -\omega_3) (\Delta +\omega_3) (\Delta +\omega_2) (\Delta -\omega_2)}}^2.
\end{align}
This expression is equivalent to the leading-order result given in Ref. \cite{Goldstein2013} for the spin-boson model. This can be understood from the fact that, at leading order, the $x\rightarrow xxx$ process does not involve any $y$ photons.
Computing the other scattering probabilities, we find that the processes $x \rightarrow \{ yyy,xxy \}$ have a simple relation to the above process, given by
\begin{dmath}
	\label{eq:gyyy}
	\gamma_{yyy}(\omega_1,\omega_2,\omega_3;\omega)=\gamma_{xxx}(\omega_1,\omega_2,\omega_3;\omega)\frac{\omega^2}{\abs{\Delta}^2},
\end{dmath}
\begin{dmath}
	\gamma_{xxy}(\omega_1,\omega_2,\omega_3;\omega)=\gamma_{xxx}(\omega_1,\omega_2,\omega_3;\omega)\frac{\omega_3^2}{\abs{\Delta}^2}.
\end{dmath}
The remaining process, $x \rightarrow xyy$, does not have a simple relation to the above expressions.
It is explicitly given by
\begin{align}
	\gamma_{xyy}(\omega_1,\omega_2,\omega_3;\omega) =&
	\frac{\alpha^4 \omega_c^4}{16}\omega_1\omega_2\omega_3\omega\abs{ \Delta }^2\times\\ \nonumber
	&\abs{\frac{ \Delta ^4+\Delta ^2 \left(\omega^2-\omega (\omega_2+\omega_3)-\omega_2^2-3 \omega_2 \omega_3-\omega_3^2\right) +\omega_2 \omega_3 (\omega_2+\omega_3)^2-\omega\omega_1(\omega_2^2-\omega_2\omega_3+\omega_3^2)}{(\omega-\Delta ) (\Delta +\omega) (\Delta -\omega_1) (\Delta +\omega_1) (\Delta -\omega_3) (\Delta +\omega_3) (\Delta +\omega_2) (\Delta -\omega_2)}}^2.
\end{align}

As with the elastic probabilities, the leading-order calculation leads to poles in the scattering probabilities at the bare spin frequency $\Delta$. We may apply procedures like those in \cref{sec:suscs} to obtain an analytic dependence, which better resembles that of the fully-interacting problem. For example, by applying the Callan-Symanzik equation to the amplitudes in Eq.~\eqref{eq:ampl1}-\eqref{eq:ampl2}, we will find that the instances of $\Delta$ will be corrected to the renormalized value $\Delta_R$. Similarly, if we developed a Dyson equation for this amplitude, we expect that the poles will be softened to broad peaks with a similar width to the peaks seen in the elastic probabilities given in the main text.

From Eq.~\eqref{eq:gyyy}, we see that scattering processes involving a photon from one waveguide into three photons in the other waveguide will dominate over scattering entirely within the same waveguide if $\omega \gg \Delta_R$. 
In this same limit, we do not expect a large region of phase space with very large final $\omega_3$, so $\gamma_{xxy}$ is expected to be much smaller than $\gamma_{yyy}$.
We have verified by numerically integrating the above expressions over the possible final frequencies that, when $\omega \gg \Delta_R$, the total cross section for the processes $x \rightarrow \{ xyy,xxy \}$ are of the same order of magnitude, and they are both much smaller than $x \rightarrow yyy$.
We also found that the cross section for $x \rightarrow xxx$ is much smaller than the three other processes in the same limit.
This is consistent with our numerical results, where we found that the inelastic scattering for $\omega \gg \Delta_R$ is dominated by scattering from one photon flavor to the other.

\section{Comparison with a model without frustration}
\label{sec:SBM1}
\begin{figure}[h]
	\centering
	\includegraphics[width=0.5\linewidth]{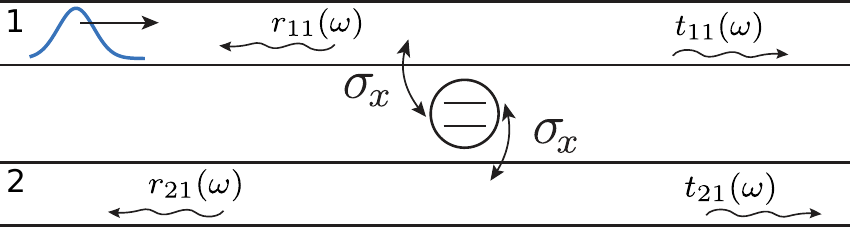}
	\caption{Schematic of the same model as in the main text but without frustration. Here the spin-1/2 is coupled locally to two independent waveguides with the same operator $\hat{\sigma}_x$.}
	\label{fig:two-waveguides-nf}
\end{figure}

In this section, we compare the results of the main text to the more common situation in waveguide QED, without frustration.
We assume the same geometry as in the main text [shown in \cref{fig:two-waveguides-nf}], with the spin coupling to two waveguides, but the coupling is via the same operator $\hat{\sigma}_x$.
The coupling constant to each waveguide is given by $\alpha_1=\alpha_2=\alpha_T/2$.

\begin{figure}[htb]
	\centering
	\includegraphics[width=\linewidth]{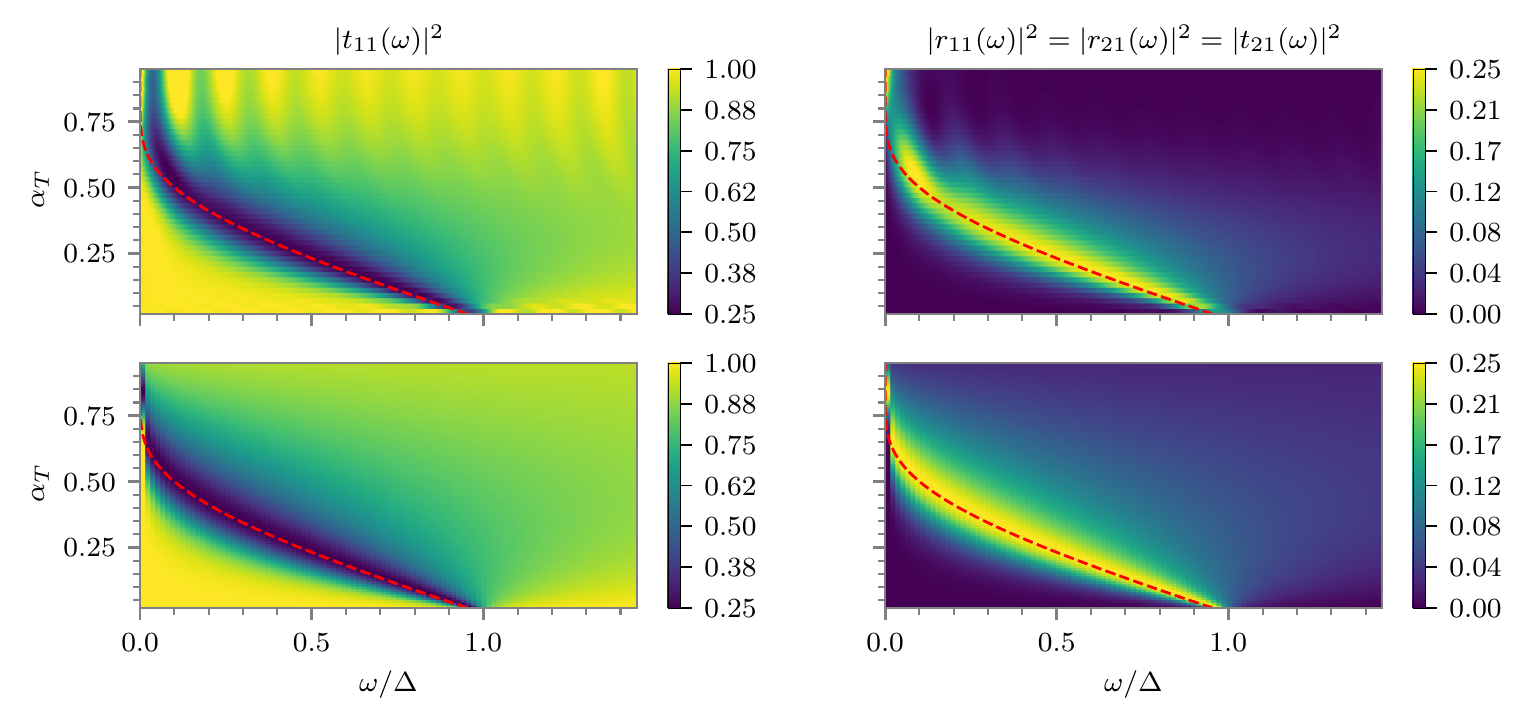}
	\caption{Numerical (top row) and analytical (bottom row) elastic scattering coefficients for \cref{fig:elasscatonebath}, as a function of the incoming frequency $\omega$ and coupling constant $\alpha_T$. The red dashed line corresponds to $\Delta_R$ from \cref{eq:Delta_R-one-bath}. The cutoff is given by $\omega_c=10\Delta$. The large oscillations at the top of the numerical $\abs{t(\omega)}^2$ are finite time/size effects due to the fact that the scattering takes a very long time at those large couplings.}
	\label{fig:elasscatonebath}
\end{figure}

We obtain the elastic scattering probabilities shown in \cref{fig:elasscatonebath} by the same method as described in the main text and in \cref{sec:numerics,sec:suscs} above. 
For the analytical calculation, the only major difference is that the $\phi_+$ and $\phi_-$ bosons are now equivalent, so the bottom two vertices in Fig.~\ref{fig:feyn_rules}(b) are equivalent to the top two.
The elastic scattering coefficients shown in \cref{fig:two-waveguides-nf} take the form
\begin{align}%
	\label{eq:reflection}
	r_{11}(\omega) & =r_{21}(\omega)=t_{21}(\omega)=-i\pi\alpha_T\omega\chi_{xx}(\omega),\\
	\label{eq:transmission}
	t_{11}(\omega) &= 1+r_{11}(\omega), 
\end{align}
and the susceptibility computed using the methods of Sec.~\ref{sec:suscs} is given by
\begin{equation}\label{eq:chixx-one-bath}
	\chi_{xx}(\omega) = \frac{-(\omega/\omega_c)^{\alpha_T}\Delta/2}{\Delta^2(\omega/\omega_c)^{2\alpha_T}-\omega^2-i\pi\alpha_T\Delta\omega(\omega/\omega_c)^{\alpha_T}}.
\end{equation}

\Cref{fig:elasscatonebath} shows that the elastic response has a resonance that is in excellent agreement with the well-known result from the spin-boson \cite{Leggett} or Kondo literature:
\begin{equation}
	\label{eq:Delta_R-one-bath}
	\Delta_R = \Delta \qty(\frac{\Delta}{\omega_c})^{\alpha_T/(1-\alpha_T)}.
\end{equation}

However, we see that, in contrast to the frustrated case, the transmission away from the resonance is very high, whereas the reflection is only nonzero around the $\Delta_R$ resonance. 
Note that $\Im(\chi_{xx}(\omega))$ in \cref{eq:chixx-one-bath}, describing the spectral weight of the spin, shows a sharp peak at $\Delta_R$.
Moreover, the large $\omega$ limit of \cref{eq:chixx-one-bath} is $\Re(\chi_{xx}(\omega)) \sim \omega^{\alpha_T-2}$ and $\Im(\chi_{xx}(\omega))\sim \omega^{2\alpha_T-3}$, both of which decay to zero much faster than in the frustrated case.
These results are consistent with previous studies \cite{Peropadre2013,Goldstein2013,Bera2016,LeHur2012,Shi2018b}, in other one-dimensional realizations of the spin-boson model.
In all of these cases, the system is well described in terms of the polaron or dressed-spin picture.
Even at very large couplings, where $\Delta_R\rightarrow0$ and the resonance is disappearing, the photons show almost no trace of the coupling to the spin, being almost fully transmitted.
This should be contrasted to the frustrated case discussed in the main text, where we have the complete opposite scenario, where the spin becomes extremely widespread over the entire energy spectrum, leading to strong elastic response in the whole range $\omega > \Delta_R$.

\begin{figure}[htb]
	\centering
	\includegraphics[width=1\linewidth]{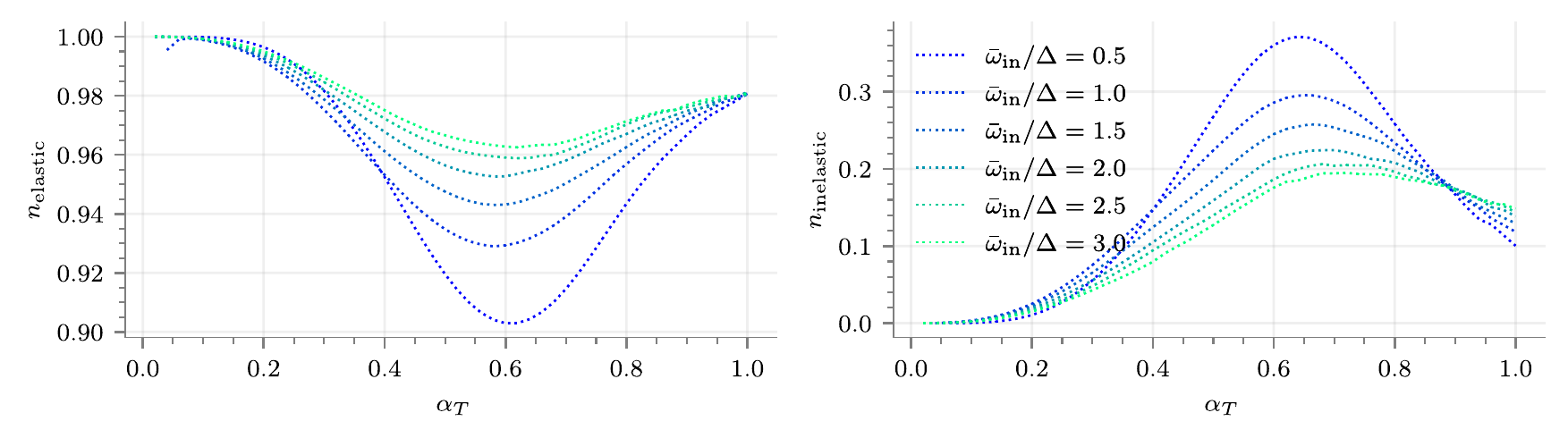}
	\caption{Numerically computed total number of elastic (left) and inelastic (right) particles as a function of $\alpha_T$ for six different incoming wavepackets, for \cref{fig:elasscatonebath}. 
		The incoming single-particle wavepackets are Gaussians centered at $\bar{\omega}_{in}$ with a standard deviation of $0.2\Delta$. The cutoff is $\omega_c=10\Delta$. }
	\label{fig:num-one-bath}
\end{figure}

Next, we look at the inelastic scattering, employing the same numerical procedure as we used in the main text. We scatter narrow wavepackets and record the resulting number of elastic and inelastic particles, shown in \cref{fig:num-one-bath}.
Note that because the coupling to both waveguides is the same, the number of inelastic photon emitted in each of the two waveguides is also the same. Hence, we present only the total number of elastic and inelastic photons.
We see that the number of elastic photons is always near $1$, never going below $\sim0.9$ for the wavepackets considered.
In fact, since $\Delta_R/\Delta \approx 0.5$ for $\alpha_T=0.25$ (see \cref{fig:elasscatonebath}), most of \cref{fig:num-one-bath} is in the regime $\omega\gg \Delta_R$. 
The number of inelastic particles does not exceed  $\sim0.36$, which occurs for the lowest-energy wavepacket $\bar{\omega}_{in}=0.5\Delta$.
In fact, as shown also in Ref.\ \cite{Goldstein2013}, for a given $\alpha_T$, the inelastic scattering rate peaks at an energy close to $\Delta_R$. For example, for $\alpha_T=0.5$ (the Toulouse point), the peak occurs at $2\Delta_R$ \cite{Goldstein2013}.
This should be contrasted again to the situation with frustration, described in the main text, where we found that the inelastic rate remained \emph{saturated} close to its maximum allowed value $0.5$ even for energies above the bar spin gap $\Delta$.

Finally, from \cref{fig:elasscatonebath} we also observe that the analytics are in much greater agreement with the numerics, even at very large $\alpha$. This is consistent with our assertion in the main text, that the disagreement in the frustrated case is due to the fact that inelastic processes are missing from the susceptibility calculation. In the frustrated case, these inelastic processes can be extremely important, accounting for half of the scattering, whereas in the present case with no frustration they are insignificant.